\documentclass[pra,showpacs,twocolumn,superscriptaddress,longbibliography]{revtex4-1}

\usepackage{graphicx}
\usepackage{color}
\usepackage{dcolumn}
\usepackage{bm}
\usepackage[caption=false]{subfig}

\usepackage{dcolumn}

\usepackage[colorlinks=true,linkcolor=red,citecolor=blue,urlcolor=blue]{hyperref}

\begin{document}

\title{Relativistic aspects of orbital and magnetic anisotropies in the chemical bonding and structure of lanthanide molecules}

\author{Eite Tiesinga}
\address{Joint Quantum Institute, National Institute of Standards
and Technology and the University of Maryland, MD 20899, USA}
\author{Jacek K\l{}os}
\address{Physics Department, Temple University,
Philadelphia, PA 19122, USA}
\address{Joint Quantum Institute, Department of Physics, University of Maryland, College Park MD 20742, USA}
\author{Ming Li}
\address{Physics Department, Temple University,
Philadelphia, PA 19122, USA}
\author{Alexander Petrov}
\address{Physics Department, Temple University,
Philadelphia, PA 19122, USA}
\address{Saint Petersburg
NRC, Kurchatov Institute PNPI, Gatchina, 188300, Russia and Division of Quantum Mechanics, Saint Petersburg University, 199034, Russia}
\author{Svetlana Kotochigova}
\address{Physics Department, Temple University, Philadelphia, PA 19122, USA}
\email{skotoch@temple.edu}

            
\begin{abstract} 
The electronic structure of magnetic lanthanide atoms is fascinating from a fundamental perspective. They have electrons in a 
submerged open 4f shell lying beneath a filled 6s shell with strong relativistic correlations leading to a large magnetic moment 
and large electronic orbital angular momentum.
This large angular momentum leads to strong anisotropies, {\it i.\,e.} orientation dependencies, in their mutual interactions. 
The long-ranged molecular anisotropies are crucial for proposals to use ultracold lanthanide atoms in spin-based quantum computers, the realization of exotic states in correlated matter, and the simulation of orbitronics found in magnetic technologies. 
Short-ranged interactions and bond formation among these atomic species have thus far not been well characterized. Efficient relativistic computations are required.
Here,  for the first time we theoretically determine the electronic and ro-vibrational states of heavy homonuclear lanthanide Er$_2$ and Tm$_2$ molecules by applying state-of-the-art  relativistic methods. In spite of the complexity of their internal structure, we were able to obtain reliable spin-orbit and correlation-induced splittings between the 91  Er$_2$ and 36  Tm$_2$ electronic potentials  dissociating to two ground-state atoms. A tensor analysis allows us to expand the potentials 
between the atoms in terms of a sum of seven spin-spin tensor operators simplifying future research.  The strengths of the  tensor operators as functions of atom separation are presented and relationships among the strengths, derived from the dispersive long-range interactions, are explained. Finally, low-lying spectroscopically relevant ro-vibrational energy levels are computed with coupled-channels calculations and analyzed. 
\end{abstract}

\maketitle

\section{Introduction}

A challenging question of molecular chemistry is an accurate description of inter-atomic and inter-molecular bonding at the 
quantum-mechanical level. This problem has attracted much attention but is not always resolved. Over the last decades, 
novel perspectives on the problem have relied on ultracold atoms and molecules. For example, quantum degenerate gases of atoms offer a unique platform on which to build and form small molecules in single internal state as they avoid unwanted 
system complexity. Ultracold gasses of atoms and molecules typically also allow for a high level of control and tunability and 
are well isolated from their surroundings. 

As part of these developments experimental breakthroughs in realizing quantum gases of atoms with large magnetic moments \cite{Pfau2005,Lahaye2009,DyBEC,ErBEC,Ferlaino2012A,Lev2012,Laburthe2013,Baier2016, Ferlaino2019} have also contributed. These atomic species
tend to have a far more complex electronic structure than that of  alkali-metal or alkaline-earth species most often studied in the field. The magnetic lanthanides from dysprosium to thulium  with their exceptionally large magnetic moments and large orbital momenta are extreme examples of such species.  This experimental research relied on controllable and tunable anisotropic dipolar interactions between the atoms. The  highly anisotropic short-range interactions between lanthanide atoms, however, remain poorly understood as they require  knowledge of their chemical bonds. These systems form an excellent environment for explorations at the interface between quantum chemistry and atomic and molecular physics.

In  previous research, we developed a successful model Hamiltonian to study the anisotropic interactions of bosonic Dy and Er  in 
an external magnetic field  \cite{Petrov2012,Kotochigova2014} and in collaboration with the experimental groups of Drs.~Ferlaino 
and Pfau we found and analyzed hundreds of magnetic Feshbach resonances in their collisions \cite{Frisch2014, Maier2015, ScienceAdv2018}. These resonances  can be used to convert an atomic gas into a gas of highly-magnetic  molecules 
as well as  to study the threshold properties or the ultracold collision-energy dependence of three-body relaxation \cite{PRL2015}. 
These atom-atom interactions have also been studied in thulium (Tm) \cite{Sukachev2010, Akimov2019}.

In spite of advances in the simulation of ultracold collisional interactions between heavy lanthanide atoms, the fundamental nature of  
the relativistic bond and short-range electronic states in lanthanide dimers as well as in the even-heavier actinide dimers remains 
mostly unexplored. Precise knowledge of these interactions is clearly desirable for predicting their quantum vibrations and 
rotations. There exists an exception though. Substantial progress  has been made in understanding interaction 
in the homonuclear diuranium molecule U$_2$ \cite{Gorokhov1974, Pepper1990, Gagliardi2005, Knecht2019}. The latest studies 
\cite{Gagliardi2005, Knecht2019} paid particular attention to the chemical bond of U$_2$ with its multi-orbital character. Relativistic 
and correlation effects using the Dirac equation for the electrons were fully incorporated by the authors of Ref.~\cite{Knecht2019} 
and enabled them to determine the energies of the lowest electronic states of U$_2$ in the vicinity of the equilibrium separation. 
In addition,  accurate  ground-state potentials for heternuclear dimer molecules that include one 
open 4f-shell lanthanide atom and one non-lanthanide atom have become available \cite{Buchachenko2007, Tomza2014, Dunning2015, Maykel2015, Tomza2018, Kosicki2020, Tomza2021}. 

The bond between two ground-state Er and two  ground-state Tm atoms is the focus  of this paper. The interactions between the 
${j=6}$  Er atoms and between the ${j=7/2}$  Tm atoms are anisotropic and orientation dependent. Here, $j$ is
the total electron angular momentum of an atom. The anisotropy is a consequence of 
potential energy differences for different  relative orientations of the electron angular momenta in the open 4f$^{12}$ and 4f$^{13}$ 
shells of Er and Tm, respectively. These 4f electrons lie beneath a closed 6s$^2$ shell so that these molecules are chemically similar but have distinct physical properties. Electron motion in lanthanides is strongly correlated and relativistic and spin-orbit coupling is strong. 

The ground-state manifold of Er$_2$ and Tm$_2$  has a large number of electronic states. They are labeled by projection quantum 
number $\Omega$ with values up to $2j$ of the total dimer electron angular momentum on the symmetry axis of the molecule 
and well as other selection quantum numbers. Because of this complexity, the intermolecular interactions until now have not been 
accurately characterized. To fulfill these objectives we have performed, for the first time,  relativistic configuration-interaction 
calculations of all $\Omega$ states as a function of interatomic separation $R$ for Er$_2$ and Tm$_2$  using the DIRAC code \cite{DIRAC19}. These configuration-interaction calculations determine the short-range energy splittings among the 91 and 36 distinct adiabatic potentials of the Er$_2$ and Tm$_2$ dimers, respectively.  

Furthermore, we have setup an analytical spin-coupling or spin-tensor representation of the short-range electronic potential surfaces for
their use in determining  rotational-vibrational levels in this paper and future improved simulations of the scattering of  ultra-cold Er 
and Tm  atoms. This representation has seven spin tensor operators and follows from the analytic form of the long-range 
anisotropic dispersion or van-der-Waals interaction. We find that the splittings among the Er$_2$ and Tm$_2$  potentials are dominated by 
a single anisotropic dipolar coupling between one of the atomic angular momenta and the mechanical rotation of the atom pair. 
We have also computed the long-range coupling strengths for the seven tensor operators based on all known atomic transition energies 
and transition dipole moments of Er and Tm. 
In fact, we find simple relationships among the seven spin-tensor operators contributing to the long-range interaction Hamiltonian. 

Finally, we predict the relativistic Hund's case (c) structure of the energetically-lowest rotational-vibrational levels of the homonuclear Er$_2$ and Tm$_2$ dimers using  a discrete-variable representation for the vibrational motion. We hope that our predictions will pave the way to spectroscopic studies of these complex and interesting molecules in the near future. 

\section{Results and discussion}

\subsection{Ground electronic states of Er$_2$ and Tm$_2$.} 

In this section we provide the relevant information on molecular electronic properties for two homonuclear lanthanide molecules, Er$_2$ and Tm$_2$, using a two-step approach to determine short-range electronic potential surfaces for all molecular states that dissociate to Er or Tm atoms in the electronic ground states [Xe]4f$^{12}$\,6s$^2({^3{\rm H}_6}$) and [Xe]4f$^{13}$\,6s$^2$($^2{\rm F}_{7/2}$), respectively. These electronic configurations contain partially-filled or open submerged 4f and chemically-active 6s atomic orbitals. Computational details and justification of the two step process are presented in Sec.~\ref{sec:methods}
as well as the Appendices. 

\begin{figure}
\includegraphics[scale=0.22, trim=0 0 0 0,clip]{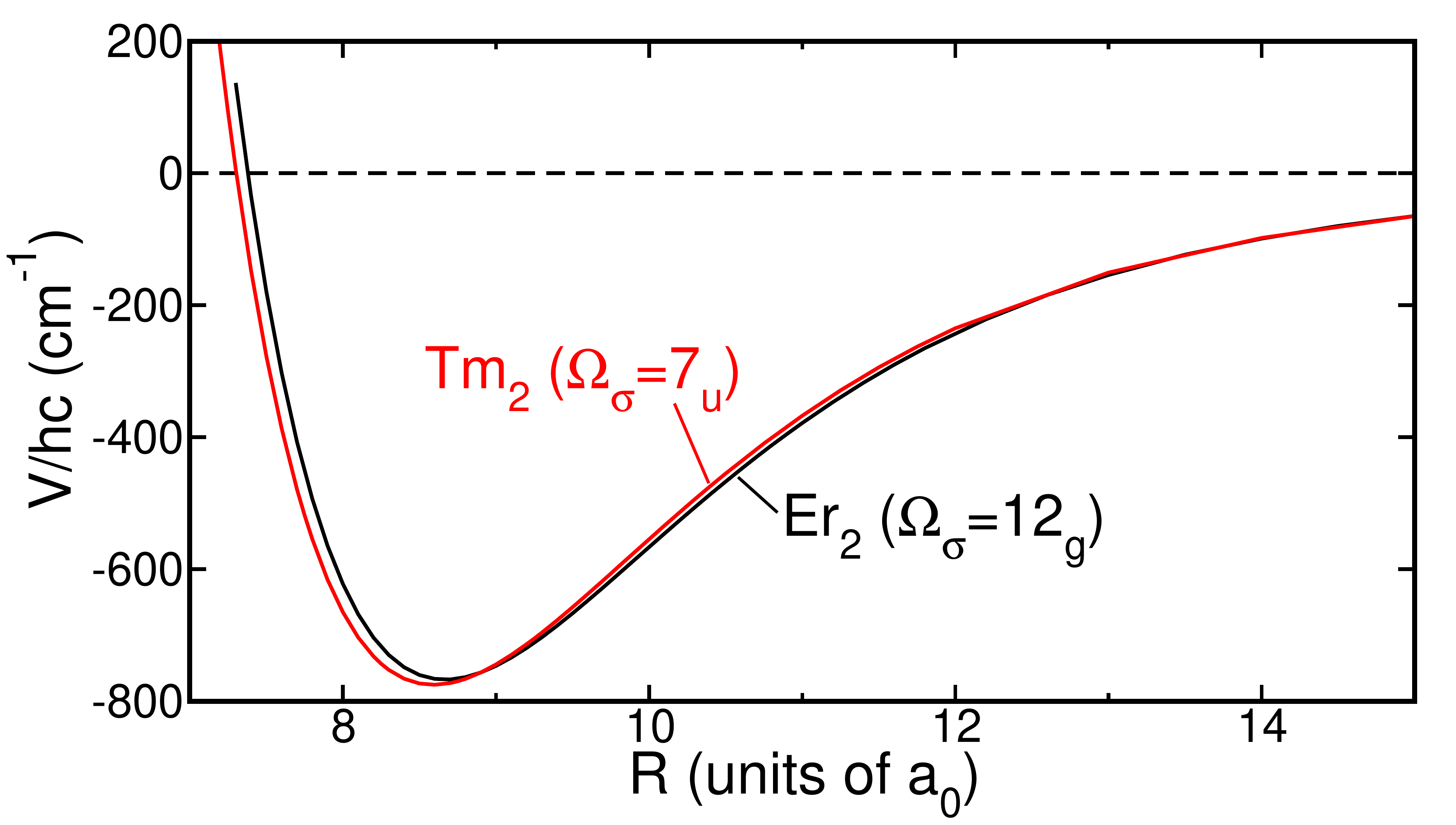}
\vspace*{-0.3cm}
\caption{Potential energies of the energetically-lowest ``spin-stretched'' states of Er$_2$ (black curve,  $\Omega_\sigma=12_g$) and Tm$_2$ (red curve, $\Omega_\sigma=7_u$) as functions of
internuclear separation $R$. The calculations are based on non-relativistic configuration-interaction and non-relativistic coupled-cluster theory for Er$_2$ and Tm$_2$, respectively. The zero of energy is at the dissociation limit of two ground-state atoms.}
\label{fig:Tm2_highspin_pec}
\end{figure}

In the first step of our study, we focus on the spin-stretched states of Er$_2$ with $\Omega_\sigma=12_g$ 
and Tm$_2$ with $\Omega_\sigma=7_u$, where subscripts $\sigma=g$ and $u$ for {\it gerade} or {\it ungerade} indicate the inversion symmetry of the electron wavefunction with respect to the center of charge.
These states have the maximum allowed total electron spin quantum number $S$ and the maximum projection quantum number $\Lambda$ of the total electron orbital angular momentum along the internuclear axis, corresponding to the $S=2, \Lambda=10$  and $S=1, \Lambda=6$ states for Er$_2$ and Tm$_2$, respectively. We have  used non-relativistic configuration-interaction or coupled-cluster calculations to determine an accurate depth for the potential energy of these spin-stretched states. 

Figure~\ref{fig:Tm2_highspin_pec} shows the spin-stretched potential energy curves for Er$_2$ and Tm$_2$ as functions of interatomic separation $R$.  The relatively shallow potential depths of just under $hc\times 800$ cm$^{-1}$ reflects the covalent bond of the two closed 6s$^2$ orbitals.  The equilibrium separations at the potential minima are $R_{\rm e}=8.7a_0$ and  $8.6a_0$ for Er$_2$ and Tm$_2$, respectively.  Here, $h$ is Planck's constant, $c$ is the speed of light in vacuum, and $a_0=0.0529177$ nm is the Bohr radius.  The depth and shape of these potentials is similar to that of the X$^1\Sigma_g^+$ state of the non-magnetic Yb$_2$ \cite{Tecmer2019}.

\begin{figure}
    \centering
    \includegraphics[scale=0.18,trim= 5 0 0 40,clip]{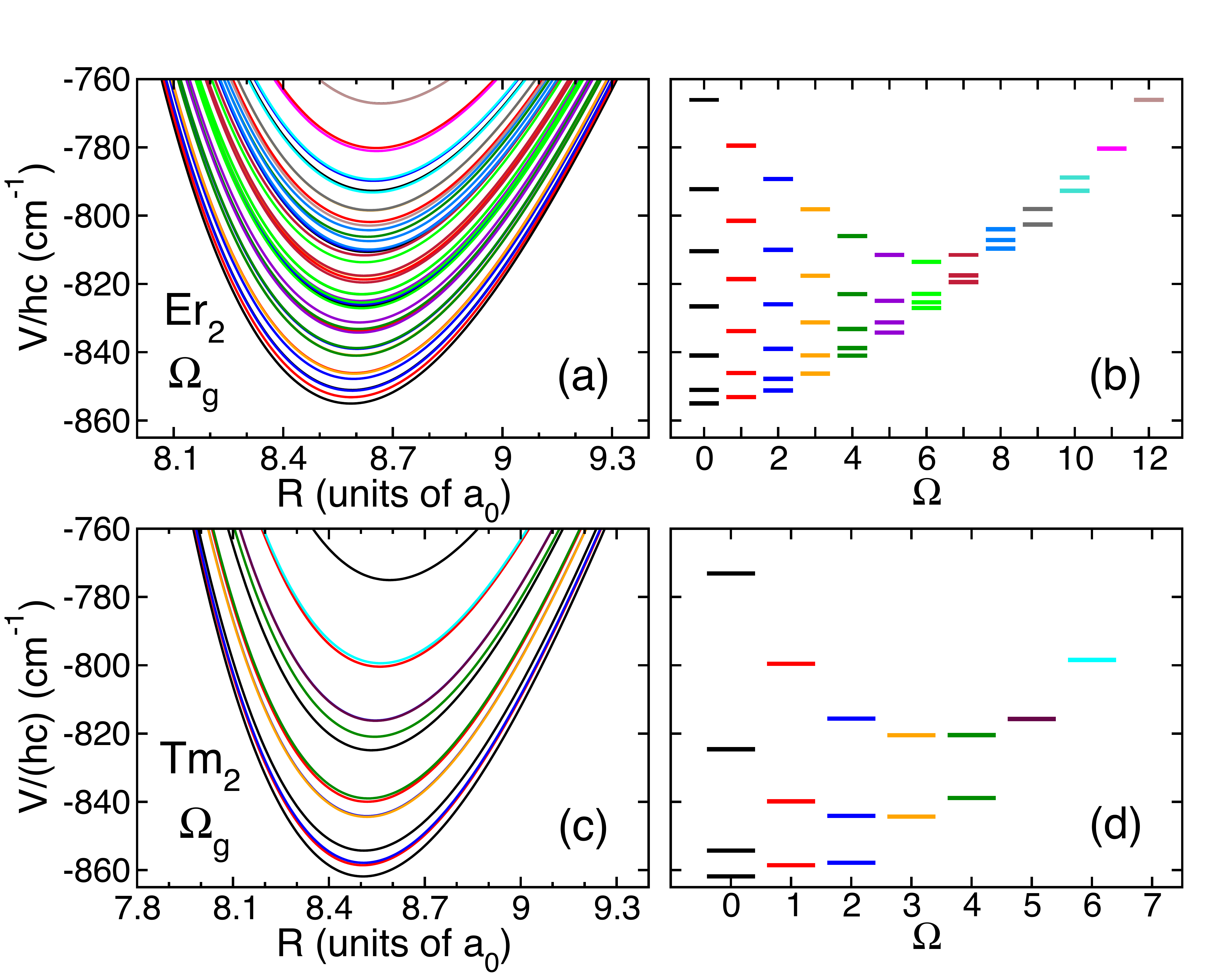}
    \caption{Relativistic $\Omega_g^{\pm}$ potential energy
curves with {\it gerade} symmetry for Er$_2$ (panel a) and Tm$_2$
(panel c) as functions of internuclear separation $R$ near the
equilibrium separation as obtained from electronic structure
calculations. All potentials approach zero energy for $R\to\infty$.
Panels (b) and (d) show potential energies from panels (a) and (c)
using the same line colors at  the equilibrium separation as
functions of projection quantum number $\Omega$ for Er$_2$ and
Tm$_2$, respectively.  {\it Gerade} $\Omega=0$ states are $0^+$
states. Potentials of states with $\Omega$ and $-\Omega$ are degenerate. }
\label{fig:Er2_Tm2_gerade_omega}
\end{figure}

In a second step, we determine energy splittings among the  Er$_2$ and Tm$_2$ potentials, using fully-relativistic configuration-interaction 
method implemented into DIRAC 2019~\cite{DIRAC19} that includes spin-orbit and anisotropic short-range interactions between atoms with 
two open 4f shells. The states are described by the projection quantum number of the total electronic angular momentum 
$\vec\jmath_{\rm el}=\vec \jmath_1+\vec \jmath_2$ on the internuclear axis $\Omega$, the {\it gerade} and {\it ungerade} symmetry, and 
a parity symmetry for $\Omega=0$ states. Here, $\Omega$ ranges from $0$ to $12$ for Er$_2$ and $0$ to $7$ for Tm$_2$ and 
labeling $\Omega^{\pm}=0^+$ or $0^-$ indicates a symmetry with respect to the reflection of the electron wavefunction through a plane containing the internuclear axis.

Figures~\ref{fig:Er2_Tm2_gerade_omega}(a) and (c) show the {\it gerade} relativistic potential energy surfaces (PES) for Er$_2$ and Tm$_2$ as functions of internuclear separation near their equilibrium separation.  There are 49 {\it gerade} potentials for Er$_2$ and 16 for Tm$_2$.  For the separations shown in the figure and, in fact, for larger separations the splittings are less than 10\,\% of the depth of the potentials relative to the dissociation energy. The figures for the {\it ungerade} states is qualitatively similar and reproduced in the Appendices. There are 42 and 20 {\it ungerade} potentials for Er$_2$ and Tm$_2$, respectively. 

The splittings among the {\it gerade} relativistic potentials seem at first glance nontrivial. A pattern, however, emerges when we plot the 
potential energies at a single $R$ near the equilibrium separation as functions of projection quantum number $\Omega$, see Figs.~\ref{fig:Er2_Tm2_gerade_omega}(b) and (d).   For both dimers the energetically lowest potential is a $0^+_g$ state. 
We also observe that the splittings among states with the same $\Omega$ gradually decrease with increasing $\Omega$.  In fact, the potential energies at the equilibrium separation are arranged in parabolic shapes.  Finally, for Er$_2$ the $0^+_g$ state with the smallest well depth is nearly degenerate with the spin stretched $12_g$ state. A discussion of the origin of this pattern is given in the following subsection. Please note that the $\Omega = 7$ state for Tm$_2$  has {\it ungerade} symmetry.

\subsection{Spin tensor decomposition of Er$_2$ and Tm$_2$ PESs.} 

A tensor decomposition of the potential energy surfaces enables us to write PESs as weighted sums of spin-spin coupling terms. It removes the need for a complicated evaluation of non-adiabatic couplings among potentials. We believe that the tensor format is essential  for our molecular systems with their tens to hundred adiabatic channel potentials in the ground configuration.  

Here, we apply the tensor decomposition technique developed in our previous study of scattering dynamics between ultracold Dy atoms 
\cite{Petrov2012} assuming that the molecular electronic wavefunction is well represented by superpositions of (anti-)symmetrized, parity-conserving products of atomic electronic ground states $| j_i m_i\rangle$ or $| j_i \Omega_i\rangle$ for atom $i=1,2$. Atomic states are labeled by eigenvalues of the total electronic angular momentum operator $\vec \jmath_i$, where projection quantum numbers along a space-fixed quantization axis are denoted by $m_i$ and those along  the body-fixed internuclear axis by $\Omega_i$. For  homonuclear systems  $j_1=j_2\equiv j$. Nevertheless, subscripts 1 and 2 on operators $\vec \jmath_i$ and
atomic states are kept to indicate the appropriate atom. (As always we omit the reduced Planck constant $\hbar$
in describing the eigenvalues of angular momentum operators.)

The atom-atom interactions are then  expressed as a sum of isotropic and anisotropic spin-tensor interactions. In principle, an infinite number of such interactions of ever increasing complexity exist.  We, however, only include the seven low-rank tensors
that describe the van-der-Waals interaction at large interatomic separations \cite{Petrov2012}. These are
\begin{equation}
		V(\vec {\mathbf R}) = \sum_{k=0,2,4}\sum_{i=1}^{N_k} V_{k}^{(i)}(R) \, \sum_{q=-k}^k (-1)^q T_{kq}^{(i)} C_{k,-q}(\hat {\mathbf R})
	\label{eq:tensor}
\end{equation}
with rank-$k$ spherical tensor operators $T_{kq}^{(i)}$ with components $q$, spherical harmonic functions $C_{kq}(\hat {\mathbf R})$ with 
${C_{kq}(\hat {\mathbf 0})=\delta_{q0}}$, and $\hat {\mathbf R}$ is the orientation of the interatomic axis. Here, $\delta_{ij}$ is the
Kronecker delta.  
The seven $T_{kq}^{(i)}$ correspond to three isotropic rank-0 tensors 
\begin{eqnarray}
	\hat{T}^{(1)}_{00} &=&I , \label{eq:I}\\
	\hat{T}^{(2)}_{00} &=&[j_1 \otimes j_2 ]_{00}/\hbar^2, \label{eq:jj0}\\
	\hat{T}^{(3)}_{00} &= &\left[ [j_1 \otimes j_1 ]_2 \otimes  [j_2 \otimes j_2 ]_2 \right]_{00}/\hbar^4 
	\,, \label{eq:qq0}
\end{eqnarray}
 three anisotropic rank-2 tensors
\begin{eqnarray}
	\hat{T}^{(1)}_{2q}& =& [j_1 \otimes j_1 ]_{2q}/\hbar^2 + [j_2 \otimes j_2 ]_{2q}/\hbar^2, \label{eq:two}\\
	\hat{T}^{(2)}_{2q} &=&[j_1 \otimes j_2 ]_{2q}/\hbar^2,  \label{eq:jj2} \\
	\hat{T}^{(3)}_{2q} &=& \left[ [j_1 \otimes j_1 ]_2 \otimes  [j_2 \otimes j_2 ]_2 \right]_{2q}/\hbar^4\,,
	\label{eq:qq2}
\end{eqnarray}
and a single anisotropic rank-4 tensor
\begin{eqnarray}
	\hat{T}^{(1)}_{4q} &=& \left[ [j_1 \otimes j_1 ]_2 \otimes  [j_2 \otimes j_2 ]_2 \right]_{4q}
	/\hbar^4\,.
	\label{eq:qq4}
\end{eqnarray}
Thus $N_0=3$, $N_2=3$, and $N_4=1$ in Eq.~(\ref{eq:tensor}). We have followed the $\otimes$ notation of
Ref.~\cite{Santra2003} for combining spherical tensor operators, which is equivalent
to the notation used in Ref.~\cite{Brink1993}.
Then $I$ is the identity operator, and $[j_1
\otimes j_2 ]_{kq}$ denotes a tensor product of angular momentum
operators $\vec{\jmath}_1$ and $\vec{\jmath}_2$ coupled to an
operator of rank $k$ and component $q$.  Finally, $V_{k}^{(i)}(R)$ are $R$-dependent strengths with
units of energy.

The eigenvalues of the interaction operator $V(\vec {\mathbf R})$ as functions of $R$ correspond to the adiabatic electronic potentials. 
The corresponding eigenstates are $R$-dependent superpositions of $| j_1 \Omega_1\rangle | j_2 \Omega_2\rangle+\epsilon\sigma | j_2 \Omega_2\rangle | j_1 \Omega_1\rangle$ (excluding normalization) with the constraints that $\Omega=\Omega_1+\Omega_2\ge 0$ and {\it gerade/ungerade} inversion symmetry are conserved. Here, ${\sigma=+1/-1}$  for {\it gerade/ungerade} states and ${\epsilon=+1/-1}$ for Er and Tm, respectively.

We obtain the strengths $V_{k}^{(i)}(R)$  by a least-squares procedure
minimizing the differences of the splittings with respect to the
spin-stretched potential.  We  only do so for separations $R\le
12a_0$ for which DIRAC 2019 converged.  All $V_{k}^{(i)}(R)$ except $V_{k=0}^{(1)}(R)$
and $V_{k=2}^{(1)}(R)$ are consistent with zero.  Thus, the dominant
spin-tensor operators are either spin independent or  a rank-2
tensor operator, corresponding to an
$R$-dependent effective atomic quadrupole moment coupled to the
rotation of the dimer.  The spin-independent strength  $V_{k=0}^{(1)}(R)$
closely follows the spin-stretch potential shown in
Fig.~\ref{fig:Tm2_highspin_pec}.  
See the Appendices for our recommended choice to construct the seven $V_{k}^{(i)}(R)$,
for uncertainty budgets, and for the procedure to construct strengths for $R>12a_0$. 

\begin{figure}
    \includegraphics[scale=0.33,trim=0 0 0 0, clip]{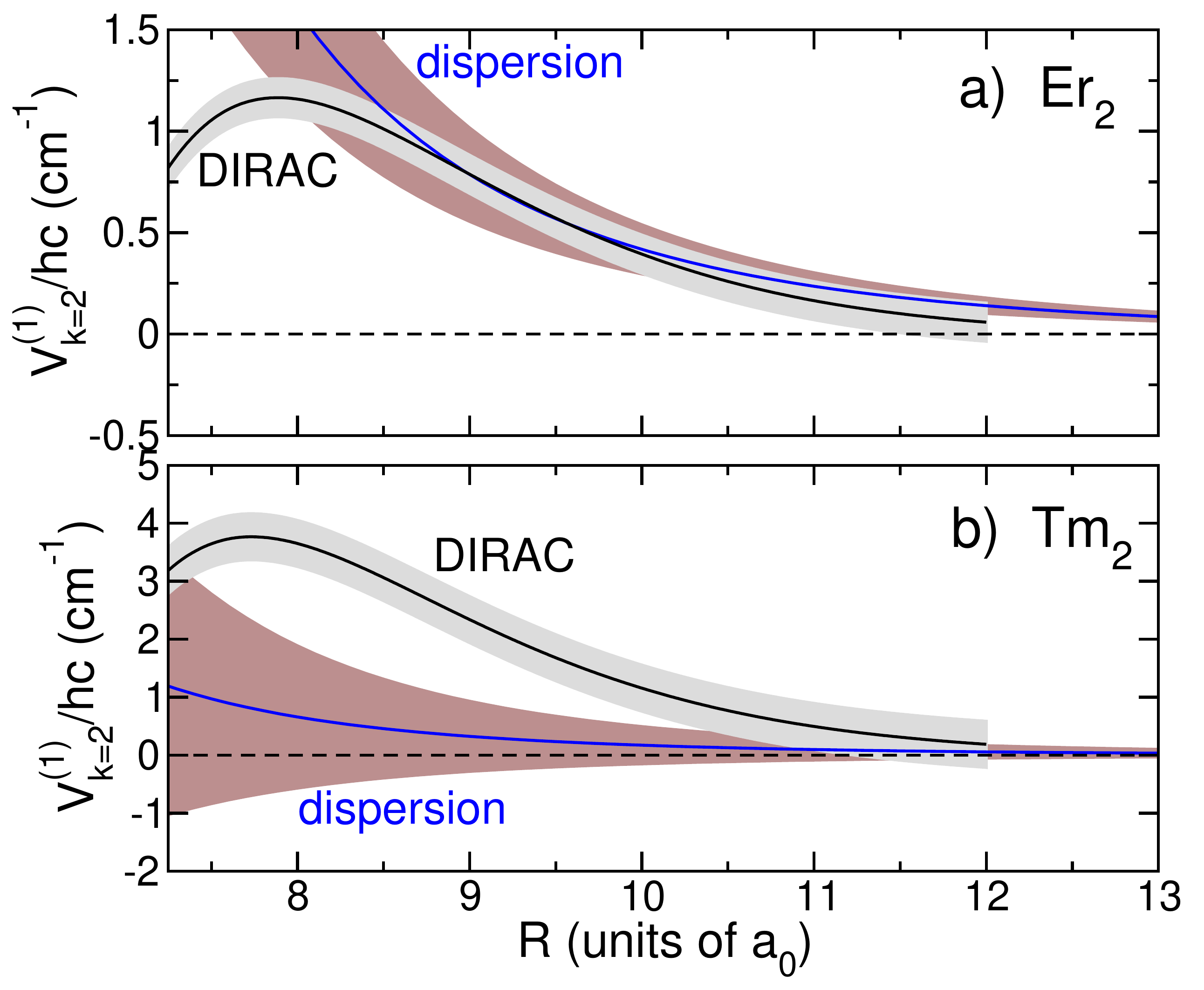}
	\caption{The fitted anisotropic spin-tensor strength $V_{k=2}^{(1)}(R)$
	for Er$_2$ (panel a) and Tm$_2$ (panel b) as a function of
	$R$ (black curve labeled DIRAC with gray one-standard-deviation
	uncertainty band).  The strength is compared to the
	van-der-Waals dispersion potential $C_{k=2}^{(1)}/R^6$ (blue
	curve with brown one-standard-deviation uncertainty band).}
	\label{fig:C16_errorbars2}
\end{figure}

The fitted $V_{k=2}^{(1)}(R)$ for Er$_2$ and Tm$_2$ are shown in Fig.~\ref{fig:C16_errorbars2}
as functions of $R$  up to $12a_0$.
The anisotropic strength is positive, is at most a few times $hc\times
1$ cm$^{-1}$, and approaches zero for large $R$. Finally, these anisotropic strengths are
at least two orders of magnitude smaller than $V_{k=0}^{(1)}(R)$.
Tables with values for $V_{k=0}^{(1)}(R)$ and $V_{k=2}^{(1)}(R)$ can be found in the Appendices.

The dominance of the isotropic spin-independent tensor operator is a consequence of the fact that the spatial extent of the 4f orbital is  much smaller than that of the 6s orbital. In fact, the 4f orbital of one atom does not significantly overlap with that of a nearby Er or Tm atom even for $R$ near the equilibrium separation. 

We can now explain the origin of the patterns seen in Fig.~\ref{fig:Er2_Tm2_gerade_omega}.
Including only the two dominant spin tensor operators in $V(\vec{\bf R})$,
 eigenstates of $V(\vec{\bf R})$  correspond to {\it single} $| j_1 \Omega_1\rangle | j_2 \Omega_2\rangle+\epsilon\sigma | j_2 \Omega_2\rangle | j_1 \Omega_1\rangle$ states. They have eigenenergies 
\begin{equation}
	 V(R;\Omega) = V_{k=0}^{(1)}(R) + V_{k=2}^{(1)}(R)
	\frac{3 (\Omega_1^2+\Omega_2^2)- 2j(j+1)}{\sqrt{6}} \,.
	\label{eq:model1}
\end{equation} 
with quadratic or parabolic dependences on $\Omega_1$ and $\Omega_2$ (and thus $\Omega$).
For Er$_2$ with positive $V_{k=2}^{(1)}(R)$ and integer  $j$,
Eq.~(\ref{eq:model1}) predicts that  the $\Omega_1=\Omega_2=0$ state and thus an
$\Omega=0$ state has the lowest potential energy.  In fact, for Er$_2$ this is
a $0_g^+$ state. In addition, Eq.~(\ref{eq:model1}) implies that the
energetically-highest $\Omega=0$ state is degenerate with the sole
spin-stretched $\Omega=12$ state. In fact, multiple degenerate
adiabatic states with the same value for $\Omega_1$  but opposite-signed
values for $\Omega_2$ exist.

For Tm$_2$ also with positive $V_{k=2}^{(1)}(R)$ but now half-integer $j$,
the model is quite satisfactory as well.  Equation (\ref{eq:model1}) predicts that states with $|\Omega_1|=|\Omega_2|=1/2$ have
the lowest energy.  In this case, $\Omega_\sigma$ is either $0_g^+$ or $0_u^-$ (both with $\Omega_1=-\Omega_2=1/2$) or $1_u$ (with $\Omega_1=\Omega_2=1/2$) and
the ground state should be three-fold degenerate. In fact, the energetically-lowest
level from the DIRAC calculations is a $0_g^+$ state. Any removal of degeneracies is  
due to one or more of the five weaker spin-tensor operators not accounted for in
Eq.~(\ref{eq:model1}).

Although the five weaker spin-tensor operators could not be reliably
extracted from the least-squares adjustment to {\it all} splittings among
the relativistic  potentials of Tm$_2$, additional analyses show that the spin-tensors in Eqs.~(\ref{eq:jj0}) and (\ref{eq:jj2})
are the most important of the five.  The first-order correction to
the energy due to these two spin-tensor operators is
\begin{equation}
	\left(-\frac{1}{\sqrt{3}} V^{(2)}_0(R)
	 +\frac{2}{\sqrt{6}}   V^{(2)}_2(R) \right) \Omega_1\Omega_2 
\end{equation}
with a positive value within the parenthesis.
Thus the $0_g^+$ state has a lower potential energy than 
the $1_u$ state.

\subsection{Spin-tensors for long-range interactions.} 

The results shown in Figs.~\ref{fig:Er2_Tm2_gerade_omega} and
\ref{fig:C16_errorbars2}  focused on the deepest parts of the
potentials.  For scattering of ultracold atoms the long-range or
large $R$ part of the potential is equally important. The long-range
form involves the van-der-Waals as well as  magnetic and quadrupolar interactions.

In our model for $V(\vec{\bf R})$ all seven strengths  $V^{(i)}_k(R)$ have a $C^{(i)}_k/R^6$
contribution for ${R\to \infty}$. Here, $C^{(i)}_k$ are van-der-Waals coefficients.  The ${k=2,i=2}$ strength has a second long-range
contribution. That is, $V^{(2)}_2(R)\to D^{(2)}_2/R^3+C^{(2)}_2/R^6$, where $D^{(2)}_2/R^3$ describes the
magnetic dipole-dipole interaction between the magnetic moments of
the lanthanides atoms.  Its strength $D^{(2)}_2$ is $-\sqrt{6}\alpha^2
(g_j/2)^2\, E_{\rm h}a_0^3$, where $g_j$ is the electronic $g$-factor
of the atomic ground state, $\alpha$ is the fine-structure
constant, and $E_{\rm h}=4.359\,74\times 10^{-18}$ J is the Hartree energy. 
The magnetic dipole-dipole interaction is not captured by electronic structure calculations, but is relevant  for scattering calculations. 

The  rank-4 strength $V^{(1)}_{k=4}(R)$ has a second long-range contribution as well. It approaches 
$Q^{(1)}_4/R^5+C^{(1)}_4/R^6$ for large $R$ with a $1/R^5$ quadrupole-quadrupole term with coefficient  $Q^{(1)}_4=6\sqrt{70}({\cal Q}/ea_0^2)^2/[j_i^2({2j_i-1})^2]\times E_{\rm h}a_0^5$
for homonuclear dimers that is solely determined by the atomic quadrupole moment ${\cal Q}=\langle j_i j_i| Q_{20}| j_i j_i\rangle$ of the ${m_i=j_i}$ spin-stretched state of  Er or Tm  \footnote{In the literature quadrupole moments are often defined to be twice ${\cal Q}$; the quadrupole operator $Q_{2q}$, however, has a unique definition. See, for example, Sec.~4.10 of Ref.~\cite{Brink1993}} and $e$ is the positive elementary charge. The quadrupole moment  for erbium was calculated in our previous paper \cite{Frisch2014} and equals $0.029 ea_0^2$.  For thulium $\cal Q$ is not available, but expected to be equally small compared to $e a_0^2$. For ro-vibrational simulations with thulium we use ${\cal Q}=0$.

We have determined the $C_{k}^{(i)}$ coefficients for Er$_2$ and Tm$_2$ from second-order perturbation theory in the electric dipole-dipole interaction using experimentally-determined atomic transition frequencies and oscillator strengths or Einstein $A$ coefficients as well as their reported uncertainties.  We closely follow the calculations
in Refs.~\cite{Petrov2012,Kotochigova2011} for the dysprosium dimer. The evaluation of the seven van-der-Waals coefficients is described in  the Appendices. Values are given in Table \ref{TmErTable}, while correlation coefficients are given in the Appendices. The relative sizes of the  $C_{k}^{(i)}$  reinforce the observations  regarding the strengths  $V^{(i)}_k(R)$ derived
from  the  electronic structure calculations

\begin{table}
\caption{\label{TmErTable} Isotropic and anisotropic van-der-Waals
dispersion coefficients $C_{k}^{(i)}$ for ${\rm Er}+{\rm Er}$ and
${\rm Tm}+{\rm Tm}$ sorted by value and then by rank $k$.  The first
three columns label the seven tensor operators. The last two
columns give their value and its one-standard-deviation statistical
uncertainties.
The strength of tensor operator $[j_1\otimes j_2]_2$ is $\sqrt{2}$ times larger than that of $[j_1\otimes j_2]_0$. Similarly, the strengths of the last three tensor operators of each dimer  are related by simple algebraic factors discussed in the text. 
For numerical
convenience the strengths are given with additional digits.
 }
\begin{tabular}{ccc|r@{.}l@{~~}r@{.}l}
\hline
\multicolumn{7}{c}{Homonuclear Erbium dimer}\\
\hline
 $k$ & $i$ &Operator $T_{k}^{(i)}$    &   
	\multicolumn{2}{c}{$C_{k}^{(i)}$}      &  
	\multicolumn{2}{c}{$u(C_{k}^{(i)})$} \\
 &     &  &     \multicolumn{4}{c}{ (units of $E_{\rm h} a_0^6$)}  
	\\
\hline
	0 &   1 & $I$                 &    $-$1723 & 072\,389\,927  &  65 & \\
	2 & 1 & $[j_1\otimes j_1]_2+[j_2\otimes j_2]_2$      &        1 & 903\,660\,883  &   0 & 57  \\[2 pt]
	0 & 2 & $[j_1\otimes j_2]_0$            &        0 & 171\,750\,953  &   0 & 099  \\
	2&2 & $[j_1\otimes j_2]_2$            &        0 & 242\,892\,527  &   0 & 14  \\[2 pt]
	0 & 3 & $[[j_1\otimes j_1]_2\otimes [j_2\otimes j_2]_2]_0$  &       $-$0 & 000\,943\,784  &   0 & 00055  \\
	2 & 3 & $[[j_1\otimes j_1]_2\otimes [j_2\otimes j_2]_2]_2$  &       $-$0 & 001\,128\,037  &   0 & 00066  \\
	4 & 1 & $[[j_1\otimes j_1]_2\otimes [j_2\otimes j_2]_2]_4$  &       $-$0 & 009\,080\,527  &   0 & 0053 \\
\hline
\\
\multicolumn{7}{c}{Homonuclear Thulium dimer}\\
\hline
 $k$ & $i$ & Operator $T_{k}^{(i)}$     & 
\multicolumn{2}{c}{$C_{k}^{(i)}$}      &
\multicolumn{2}{c}{$u(C_{k}^{(i)})$} \\
	&     &         &     \multicolumn{4}{c}{ (units of $E_{\rm h} a_0^6$)}  
	\\
\hline
	0&    1 & $I$                 &    $-$1672 & 115\,030\,649  &  54 &   \\
	2 & 1 & $[j_1\otimes j_1]_2 +[j_2\otimes j_2]_2$     &        0 & 788\,488\,761  &   1 & 47  \\[2 pt]
	0 & 2 & $[j_1\otimes j_2]_0 $           &        0 & 001\,566\,976  &   0 & 012  \\
	2& 2	 & $[j_1\otimes j_2]_2 $           &        0 & 002\,216\,039  &   0 & 017  \\[2 pt]
	0 & 3 & $[[j_1\otimes j_1]_2\otimes [j_2\otimes j_2]_2]_0$  &       $-$0 & 000\,309\,025  &   0 & 00060  \\
	2 & 3 & $[[j_1\otimes j_1]_2\otimes [j_2\otimes j_2]_2]_2$  &       $-$0 & 000\,369\,355  &   0 & 00072  \\
	4 & 1 & $[[j_1\otimes j_1]_2\otimes [j_2\otimes j_2]_2]_4$  &       $-$0 & 002\,973\,250  &   0 & 0058\\
		\hline
\end{tabular}
\end{table}

For Er$_2$ we  observe that the absolute value of the magnetic dipole-dipole interaction $|D_2^{(2)}|/R^3$ equals $C_2^{(1)}/R^6$ at $R\approx35a_0$ and $C_2^{(2)}/R^6$ at $R\approx18a_0$. Hence, for $R\gg 35a_0$ the magnetic dipole-dipole interaction is the strongest anisotropic interaction, while for smaller $R$ the effective quadrupole interaction of Eq.~(\ref{eq:two}) is the strongest anisotropic interaction.

Less obvious from  Table \ref{TmErTable} is that we have
been able to derive non-trivial algebraic relationships among the $C^{(i)}_k$
thereby reducing the number of independent dispersion coefficients from 7
to 4. We find that
\begin{equation}
   C_2^{(2)}= \sqrt{2}\, C_0^{(2)} \label{eq:c22}
\end{equation}
showing that the spin-exchange  strength and the effective dipole-dipole strength multiplying Eqs.~(\ref{eq:jj0}) and (\ref{eq:jj2}), respectively, are related.
Similarly, we find that
\begin{equation}
   C_{2}^{(3)}= \sqrt{\frac{10}{7}} C_{0}^{(3)}\quad {\rm and} \quad
   C_{4}^{(1)}= 6\sqrt{\frac{18}{7}} C_{0}^{(3)}
\end{equation}
relating the strengths of the three spin-tensors constructed from the two
effective atomic quadrupole operators 
$[j_1 \otimes j_1 ]_2$ and $[j_2 \otimes j_2 ]_2$.
The derivation of these relations can be found in the Appendices.

\subsection{Ro-vibrational eigenstates.} 

We finish our analyses of Er$_2$ and Tm$_2$ by computing their energetically-lowest ro-vibrational eigenstates. That is, we  compute  eigenstates of 
$-\hbar^2\nabla^2/(2\mu)+V(\vec {\bf R})$, where   $\mu=m/2$ and
$m$ is the mass of the Er or Tm atom
\footnote{In this section we have implicitly made a change in
coordinate system, where ${\bf R}$ is now the separation between the
center of masses of the atoms rather than the separation between the nuclei.
Similarly, we use  atomic rather than nuclear masses in the kinetic energy
operator. These ``non-adiabatic'' changes are not considered significant in light of the 
uncertainties of our interaction potentials.}. We discretize the radial component of the kinetic energy operator $-\hbar^2\nabla^2/(2\mu)$ using the discrete-variable representation of Ref.~\cite{Colbert1992}. Details regarding the selection of the spin and angular momentum basis and, in particular,
the orbital or partial-wave angular momentum $\vec \ell$ and total molecular angular momentum $\vec J$ of the two rotating atoms are given in Sec..
We  present results from two calculations. One is based on potential $V(\vec{\bf R})$ including only the two
dominant spin tensors ${\hat T}_0^{(1)}$ and ${\hat T}_2^{(1)}$, constructed by joining the electronic structure data with the long-range
potentials,  and one where all seven tensor operators are included. For latter case, the $R$-dependence of the remaining five weaker spin tensor operators is given by their long-range form for all $R$.

\begin{figure}
    \includegraphics[scale=0.31,trim=0 0 0 0, clip]{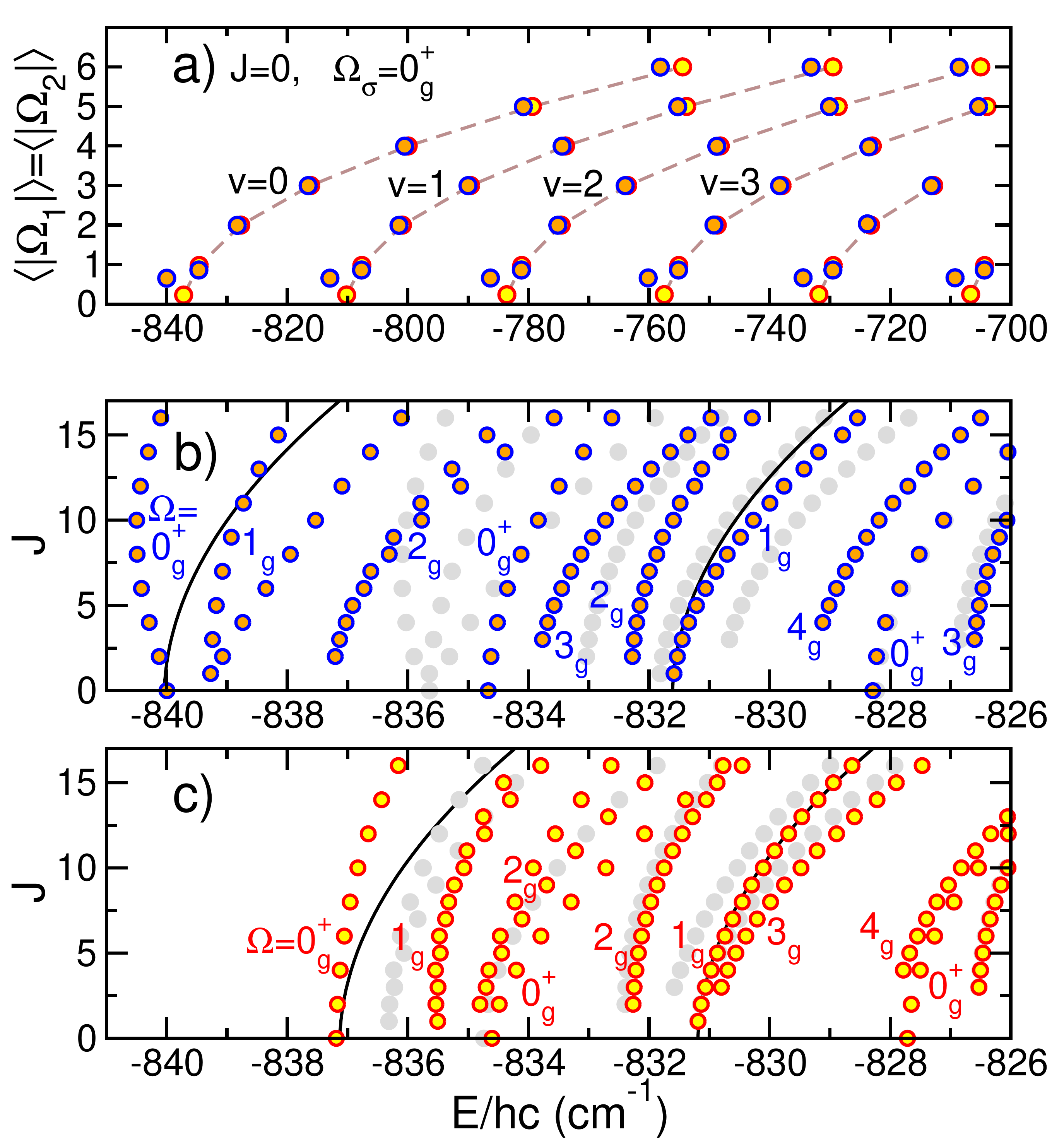}
\caption{Three views of eigenenergies $E$ of  bosonic $^{168}$Er$_2$  near the minima 
of its potentials. Energies  are with respect to the dissociation limit of two 
$^{168}$Er atoms. Panel a) shows the energies of  
the fine-structure components for the six energetically-lowest  $J=0$ vibrational levels $v$ of
the $0_g^+$ state. For each $v$ these $0_g^+$ states
are uniquely labeled by the expectation value of  $|\Omega_1|$, which are close to integer valued.
Orange-filled and yellow-filled circles are the result of calculations that include all spin-tensor interactions and only include the two strongest spin tensors, respectively.  
Panels b) shows the energies of  $v=0$ rotational eigenstates versus $J$ for  calculations that include all tensor operators,
while panel c) shows equivalent data including only the two strongest spin tensors. 
Labeled colored  circles  correspond to {\it gerade} states.
Solid gray markers correspond to the energies of  {\it ungerade} states.
Finally,    black curves in  panels b) and c) correspond to   
progressions $B_{\rm e}J(J+1)$, where $B_{\rm e}=hc\times 0.0095$ cm$^{-1}$ is the rotational constant at the equilibrium separation.
}
	\label{fig:ErLevels}
\end{figure}

Figure \ref{fig:ErLevels} shows three views of ro-vibrational eigenenergies of bosonic $^{168}$Er$_2$ states near the minimum of the adiabatic potentials. As the nuclear spin of  $^{168}$Er  is zero  {\it gerade} basis states have  even values for partial wave $\ell$.  {\it Ungerade} basis states require odd  $\ell$. 
The pattern of the  energy levels in the $hc\times 150$ cm$^{-1}$ energy range  in Fig.~\ref{fig:ErLevels}a) can be understood from the seven $0_g^+$ adiabatic potentials shown in Figs.~\ref{fig:Er2_Tm2_gerade_omega}a) and b) and Eq.~(\ref{eq:model1}). The vibrational energy spacing based on the harmonic approximation around the minima of the nearly-parallel adiabatic potentials is $ hc\times 27.0$ cm$^{-1}$ so that vibrational levels $v=0,1, \cdots, 5$ are visible in the figure. For each $v$ the spacings among the seven $0_g^+$ states are to good approximation found from $\sqrt{6}  V_{k=2}^{(1)}(R_{\rm e})\Omega^2_1$ for $|\Omega_1|=0,\cdots,6$ with $\sqrt{6}V_{k=2}^{(1)}(R_{\rm e})=hc\times 2.3$ cm$^{-1}$ at the equilibrium separation. The $\Omega_1$ fine-structure thus overlaps with the vibrational structure.

Figure \ref{fig:ErLevels}b)  shows  ${v=0}$ $\Omega_{g/u}^\pm$ $^{168}$Er$_2$ eigenstates
over an energy region of only $hc\times 15$ cm$^{-1}$  versus total molecular angular momentum $J$. 
based on calculations that include all spin tensors.
Panel c) shows equivalent data including only the two strongest spin tensors.
The level density in both  two panels is large, although the level patterns are distinct with differences around $hc\times 1$ cm$^{-1}$. 
In other words,  the weaker spin-tensors can not be ignored in an accurate analysis of the lowest energy states
of Er$_2$.

Surprisingly, we predict that the $J=10$ rotational state of the 
$v=0$ $0_g^+$ ground state has the absolute lowest energy when all interactions are included.
The reason for this and other unexpected rotational progressions
is the degeneracies of different $\Omega$ states inherent in the model of Eq.~(\ref{eq:model1})
as well as more accidental degeneracies due to the high level density. The coriolis force in a rotating molecule breaks
these (near-)degeneracies for states with $\Omega$ values that differ by one unit. For large $J$ coupling
matrix elements are on the order of $B_{\rm e} J j_1$, which can easily reach
values of order $hc \times 1$ cm$^{-1}$ comparable to or larger than $ V_{k=2}^{(1)}(R_{\rm e})$, even when $B_{\rm e}$ is not.
Degenerate perturbation theory  then predicts ``rotational'' progressions
of the form $\pm A_{\rm e} J + B_{\rm e}J(J+1)$, where energy $A_{\rm e}$ with $|A_{\rm e}|\gg B_{\rm e}$ can be 
computed on a case by case basis.

\begin{figure}
    \includegraphics[scale=0.32,trim=10 20 5 5, clip]{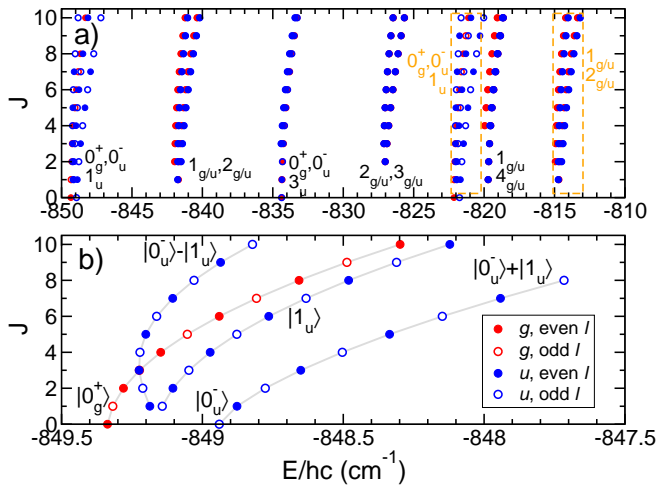}
\caption{Eigenenergies $E$ of  $^{169}$Tm$_2$ near the minima 
of its potentials. Energies  are with respect to the dissociation limit of two 
$^{169}$Tm atoms. Panel a)  shows energies of  $v=0$ and $v=1$ $\Omega_\sigma^\pm$ 
vibrational and fine-structure eigenstates versus total molecular angular momentum 
$J$ for  calculations that include all spin-tensors interactions. Levels enclosed 
by orange dashed boxes are $v=1$ states. All others are $v=0$ states.
Panel b) shows a blowup of the energies of the energetically-lowest $0_g^+$, $0_u^-$, and 
$1_u$ states. Unusual rotational progressions are visible and are discussed in the text.
Color and type of markers, as indicated in the legend box, separate {\it gerade} from 
{\it ungerade} states as well as even from odd partial wave $\ell$ states and are the 
same in both panels.
}
	\label{fig:TmLevels}
\end{figure}

Figure \ref{fig:TmLevels} shows the lowest  eigenenergies of  $^{169}$Tm$_2$ 
near the minimum of its adiabatic potentials versus total molecular angular momentum $J$. 
Both {\it gerade} and {\it ungerade} states are shown and assigned $\Omega_{g/u}^\pm$ labels.
The nuclear spin $\vec \imath$ of  $^{169}$Tm is 1/2, making the atoms bosons, and hyperfine 
interactions of the form $a_{\rm hf} \vec \jmath_1\cdot \vec \imath_1/\hbar^2$ etc. mix {\it gerade} 
and {\it ungerade} states, although even and odd $\ell$ remain uncoupled. 
We do not include such interactions and {\it gerade} states with either even and 
odd $\ell$ must be shown. Similarly, {\it ungerade} states with  even and odd $\ell$ exist.
It is worth noting that the $^{169}$Tm   hyperfine constant 
$a_{\rm hf}=hc\times0.0062$ cm$^{-1}$ \cite{Sukachev2010}
and hyperfine couplings can mostly be neglected in the analysis of the bound states on the scale shown in the figure.

The Tm$_2$ level structure is significantly simpler than that for Er$_2$, even though  
the vibrational spacings and rotational constants of the two dimers are nearly the same.
The dominant anisotropic spin-tensor interaction in Tm$_2$ is close to three times larger than in Er$_2$, as seen in
Fig.~\ref{fig:C16_errorbars2}, thereby reducing the number of $\Omega_\sigma^\pm$ states
between the $v=0$ and 1 vibrational states.  
The vibrational and rotational   spectroscopic constants for Tm$_2$ are  $hc\times 27.2$ cm$^{-1}$ and $hc\times 0.0096$  cm$^{-1}$, respectively. Unlike for Er$_2$ the remaining five weaker spin-tensor operator play no significant role.

Figure~\ref{fig:TmLevels}b) shows that  the $v=0$, $J=0$ $0_g^+$ 
state is the absolute ground state. This $0_g^+$ state has a ``simple'' $B_{\rm e}J(J+1)$
rotational progression.
The nearby {\it ungerade} states have a less-conventional rotational 
progression. Here, this follows from the {\it four}-fold degeneracy for the lowest-energy states implied by 
Eq.~(\ref{eq:model1}). The four states 
have $|\Omega_1|=|\Omega_2|=1/2$ and  labels $\Omega_\sigma^\pm=0_g^+$, $0_u^-$ and twice $1_u$. 
In fact, a careful analytical analysis of the centrifugal and coriolis interactions within the degenerate
manifold shows that in the limit $J\to 0$
the {\it ungerade} levels have energies that lie tens of $B_{\rm e}$
 above that of  the $0_g^+$ state. 
For large $J$ two of the three {\it ungerade} states become equal mixtures
of $1_u$ and $0^-_u$.
The {\it ungerade} state with the second-lowest energy remains of pure $1_u$ symmetry.
Finally, we observe that the energetically-lowest {\it ungerade} state has ${J=3}$.

\section{Methods}\label{sec:methods}

\subsection{Spin-stretched electronic potentials.} 

The spin-stretched  potentials $V_{\rm ss}(R)$
for Er$_2$ and Tm$_2$ have been obtained from (par\-tial\-ly-) spin-restricted single-reference
non-relativisitic coupled-cluster calculations that included single, double and perturbative triple (RCCSD(T)) excitations \cite{Knowles1993}.
For the calculations the total electron spin $S$ is 2 and 1 for Er$_2$ and Tm$_2$, respectively. In addition,
the  projection quantum number $\Lambda$ of the total electron orbital angular momentum along the internuclear axis
is 10  and 6 for Er$_2$ and Tm$_2$, respectively. 

We use the Stuttgart/Cologne ``small-core'' quasi-relativistic effective
core potentials (ECPs) developed for rare-earth elements
(ECP28MWBSO)\cite{Dolg1989} to describe the twenty eight 1s, 2sp,
and 3spd electrons of the Er and Tm atoms. The remaining electrons
are described by the Relativistic Small Core Segmented
(RSCSEG)\cite{Cao2002} atomic basis set of quadruple-zeta quality
developed for the ECPs. We extend the basis set with three diffuse Gaussian s functions
with  exponents 0.1495, 0.01, and 0.00412; two p functions with
exponents 0.04895 and 0.0211; one d function with exponent 0.02799;
and one f and g function both with exponent 0.1068, respectively.
All exponents are in units of $a_0^{-2}$.  In order to converge the preliminary self-consistent-field (SCF) calculations of neutral Er$_2$ and Tm$_2$, 
we start from the SCF orbitals for molecular ions Er$_2^{4+}$
and Tm$_2^{4+}$. 

Based on a comparison of results found with different basis set size,
the one-standard-deviation uncertainty of the
spin-stretched potential is about $hc\times 50$ cm$^{-1}$ at the equilibrium
separation and drops to less than $hc\times 1$ cm$^{-1}$ at $R=20a_0$.  See
tables  in the Appendices for precise data on the spin-stretched potentials.
A description of the extrapolation to $R>20a_0$ can also be found there.

\subsection{Relativistic calculation of potential splittings.} 

We use the  direct relativistic configuration interaction (DIRRCI)
method~\cite{VISSCHER1994120} in DIRAC \cite{DIRAC19} to determine
the energy splittings between the relativistic adiabatic potential
curves of Er$_2$ and Tm$_2$ dissociating to two  ground-state atoms. 

As in the non-relativistic calculations converging SCF calculations
for the neutral dimers start from SCF orbitals for Er$_2^{4+}$
and Tm$_2^{4+}$. A reordering of the occupied 6s and
open-shell 4f orbitals is then required in the input data for DIRAC.
In practice, we only determine the SCF orbitals at $R=12a_0$ in this manner.  
SCF orbitals  for $R<12 a_0$ are  found starting from orbitals for the
neutral dimer obtained for a slightly larger $R$.  We
repeat the scheme to small $R$ until the potentials
are repulsive.

The active space in the DIRRCI calculations  is solely
composed of molecular orbitals arising from the 4f atomic shells.
The 6s orbitals are kept doubly occupied and are not part of the
active space. In addition, 5d orbitals remain unoccupied.  These
constraints balance the need for reasonable estimates of splittings
among the relativistic potentials and a reasonable run-time and memory usage for the calculations.  
For state assignment it proved useful to determine the expectation values of the $z$-components
of the total electronic orbital angular momentum and spin operators along the internuclear axis.

DIRRCI calculations have been performed for 25 and 28 internuclear separations between $7a_0 < R\le 12a_0$
for Er$_2$ and Tm$_2$, respectively.  
At each of these $R$ values, we  determine the 91 relativistic potential energies and eigenstates for Er$_2$.
For Tm$_2$ we find 36 eigenpairs.
An eigenstate is uniquely labeled by  $n,\Omega^{\pm}_{g/u}$ with $n=1,2,3,\ldots$ for  $\Omega^{\pm}_{g/u}$ states ordered by
increasing potential energy. 
We denote relativistic eigenenergies by $U_{\rm rel}(R;n,\Omega^{\pm}_{g/u})$.
For the $12_g$ and $7_u$  spin-stretched state for Er$_2$ and Tm$_2$, respectively, there
exists just one potential dissociating to two ground state atoms.
 For our identical ground-state atoms {\it gerade} states are $0^+$ states, while {\it ungerade} states are $0^-$ states.  A discussion of di-atomic symmetries and molecular state labels can be found in Refs.~\cite{Herzberg1950} and \cite{Hougen1970}.  

With the constraints on the active space in the DIRRCI calculations,
we sacrificed on the accuracy of the depth of the potentials. Those are mainly determined 
by excitations of electrons out of the 6s orbitals.  In order to obtain accurate adiabatic
potential energy curves  $V_{\rm rel}(R;n,\Omega^{\pm}_{g/u})$, we  assume that the  non-relativistic
spin-stretched potential $V_{\rm ss}(R)$ is a good representation of the potential for the relativistic
spin-stretched state. 
The adiabatic potentials for the other adiabatic states are then
\begin{eqnarray}
V_{\rm rel}(R;n,\Omega^{\pm}_{g/u})&=& V_{\rm ss}(R) \label{eq:rela1}
       \\
       &&
      - \left(U_{\rm rel}(R;n,\Omega^{\pm}_{g/u})-U_{\rm rel}(R;1,\Omega_{\rm ss})\right)\,\
 \nonumber
\end{eqnarray}
when $R\le 12a_0$ and  $\Omega_{\rm ss}=12_g$ and 7$_u$ for Er$_2$ and Tm$_2$, respectively. 
The uncertainties in the $R\le 12a_0$ calculations and extrapolation to $R>12a_0$ using the long-range dispersion potentials are discussed in the Appendices. The potentials in Eq.~(\ref{eq:rela1}) are used in the least-squares fitting to spin-tensor operators as  described in the main text.

\subsection{Basis sets in ro-vibrational state calculations.} 

We use the unit-normalized coupled spin and angular momentum basis
\begin{equation}
 |(j_{\rm el}\ell)J M\rangle \equiv
      \!   \sum_{m_j m_\ell} \!
            | (j_1j_2)j_{\rm el} m_{\rm el} \rangle Y_{\ell m_\ell}({\bf \hat R})
            \langle j_{\rm el}  \ell \, m_{\rm el} m_\ell | J M \rangle
\end{equation}
for the calculation of  ro-vibrational states of Er$_2$ and Tm$_2$ with 
\[ | (j_1j_2)j_{\rm el}  m_{\rm el} \rangle=\sum_{m_1m_2} |j_1 m_1\rangle |j_2 m_2\rangle
 \langle j_1 j_2 m_1 m_2 | j_{\rm el}  m_{\rm el}\rangle
 \]
and spherical harmonic functions $Y_{\ell m}({\bf \hat R})$. Here, $\langle j_1 j_2 m_1 m_2 | j m\rangle$ are Clebsch-Gordan coefficients. The total molecular angular momentum 
$\vec J=\vec \ell + \vec \jmath_{\rm el} $ is conserved, $\vec \jmath_{\rm el}  =\vec \jmath_1+\vec \jmath_2$, and projection quantum numbers $m_x$ and $M$ are with respect to  a space-fixed coordinate system.  
Basis states with even and odd $j_{\rm el}$ contribute to
{\it gerade} and {\it ungerade} molecular states, respectively, and
are not mixed by the molecular Hamiltonian.
Similarly, the Hamiltonian does not couple basis states  with even partial wave $\ell$ with those with odd $\ell$.
Atomic masses have been taken from Refs.~\cite{Huang2017,Wang2017} and
atomic $g$-factors from Ref.~\cite{Kramida2019}. 

\section{Conclusions}

We have studied the electronic properties of two heavy homonuclear lanthanide molecules, Er$_2$ and Tm$_2$.
A hybrid non-relativistic/relativistic electronic structure approach was needed to overcome the computational challenges
arising from the  complexity of their open submerged 4f electronic shell structure partially hidden by a closed  6s$^2$ shell.
This allowed us for the first time to determine a complete set of ground-state potentials for a wide range of interatomic separations. 

A non-relativistic coupled-cluster calculation was  used to determine the spin-stretched potential energy surfaces with the maximum allowed total electron spin $S$ and projection quantum number $\Lambda$ of the total electron orbital angular momentum along the internuclear axis. 
Then we used a relativistic multi-configuration-interaction calculation to determine the splittings among the potentials dissociating to two ground state atoms. There are 91 {\it gerade/ungerade} potentials for Er$_2$ (with $\Omega$s from 0 to 12)  and 36 potentials for Tm$_2$ (with $\Omega$s from 0 to 7). We identified the splittings as due to  different relative orientations of the angular momenta of 4f shell electrons. 

To facilitate the application of our electronic structure predictions in spectroscopic and scattering dynamics studies we analytically expressed the potential energy operator for Er$_2$ and Tm$_2$ as a sum of a small number of spherical-tensor operators and elucidated the relationships between their electrostatic, relativistic, and magnetic dipole-dipole interactions. The most remarkable aspect of  this analysis is that to good approximation the potential energy operator can  be  described with only two spin-tensor interactions, one isotropic and one anisotropic. 

Finally, we computed the spectroscopically relevant lowest ro-vibrational eigenstates of Er$_2$ and Tm$_2$. This data can be used as  preliminary 
information for setting up spectroscopic studies of these exotic and technologically important systems.

\vspace*{0.5cm}
\noindent
{\bf \large{Acknowledgements}}\\
\noindent
Work at Temple University is supported by the U.S. Air Force Office of Scientific Research Grant \#FA9550-21-1-0153, the Army Research Office Grant \#W911NF-17-1-0563, and the NSF Grant \#PHY-1908634.

\vspace*{0.5cm}
\noindent{\bf \large Data availability statement}\\
\noindent
The data in this paper is self contained.

\appendix
\section{Table of content for Appendices}

These appendices contain the input data and a derivation needed to reproduce the
potentials presented in our article on the interactions between
two ground-state erbium atoms and two ground-state thulium atoms.
We give tables for non-relativistic spin-stretched potentials and relativistic anisotropic spin-tensor
strengths. The main text has a graph of the potentials of {\it gerade} states of the two dimers.
Here, we present the equivalent figure of potentials for {\it ungerade} states.

In addition, we derive the long-range van-der-Waals dispersion interactions
for our high-spin Er and Tm atoms. We find that these interactions can
be described in terms of seven spin-tensor operators, whose strengths or van-der-Waals coefficients
are linearly dependent. In fact, only four independent dispersion coefficients exist.
We also give tables of Er and Tm atomic transition frequencies and Einstein $A$
coefficients or oscillator strengths  on which the values of the  dispersion coefficients 
in the table in the main text are based. 

We use Planck's constant $h$ and the speed of light in vacuum  $c$ in converting energies into wavenumbers.

\section{Non-relativistic spin-stretched states of Er$_2$ and Tm$_2$}\label{app:nonrel}

The spin-stretched  potentials for Er$_2$ and Tm$_2$ dissociating to two ground-state atoms have been obtained from 
non-relativistic single-reference coupled-cluster calculations that included single, double and perturbative triple excitations (RCCSD(T)). 
The orbital basis sets for these calculations have been described in Sec.~\ref{sec:methods}.
The total electron spin $S$ and orbital-angular-momentum projection quantum number $\Lambda$ are conserved quantities
and for this state have their 
maximum allowed value. We have $(S,\Lambda)=(2,10)$ and $(1,6)$ for Er$_2$ and Tm$_2$, respectively. 

The spin-stretched potential $U_{\rm ss}(R)$ for Er$_2$ has been determined with coupled-cluster
theory as implemented in CFOUR \cite{CFOUR} at 59 separations between
${R_{\rm min}=7.3a_0}$ and ${R_{\rm max}=20a_0}$ as well as at
$R_\infty=200a_0$ in order to determine the dissociation energy of
the potential.  Here, $a_0 = 0.0529177$ nm is the Bohr radius.
The spin-stretched potential for Tm$_2$ has been
determined using coupled-cluster theory as implemented in Molpro
\cite{MOLPRO} at 72 separations between ${R_{\rm min}=6.25a_0}$ and
${R_{\rm max}=20a_0}$ as well as at $R_\infty=60a_0$.
Potentials 
\begin{equation}
V_{\rm ss}(R)=U_{\rm ss}(R)-U_{\rm ss}(R_\infty)
\end{equation} 
of Er$_2$ and Tm$_2$ up to $R=R_{\rm max}$ are given in Tables \ref{tab:ErSS} and \ref{tab:TmSS}, respectively.

\begin{table*}
\caption{Potential energy $V_{\rm ss}(R)$ of the energetically-lowest
``spin-stretched'' state of Er$_2$ $\Omega$=12$_g$ as function of
internuclear separation $R$. The $R$-dependent uncertainty of this potential is discussed in the text.}
	\label{tab:ErSS}
	\begin{tabular}{c@{~}c@{~~}|@{~}c@{~}c@{~~}|@{~}c@{~}c}
	$R/a_0$  &   $V_{\rm ss}/hc$ (cm$^{-1}$) &
		$R/a_0$  &   $V_{\rm ss}/hc$ (cm$^{-1}$) &
		$R/a_0$  &   $V_{\rm ss}/hc$ (cm$^{-1}$) 
	\\
\hline
              7.3       &   136.928967055564    &               9.3       &  $-$702.972094884141    &              11.6       &  $-$290.170412396272   \\
              7.4       &  $-$33.4063015426345    &               9.4       &  $-$685.377674371753    &              11.8       &  $-$265.152758759965   \\
              7.5       &  $-$179.030466142739    &               9.5       &  $-$666.764780470275    &              12.2       &  $-$221.189686201221   \\
              7.6       &  $-$302.624146371783    &               9.6       &  $-$647.392883509601    &              12.6       &  $-$184.510652309416   \\
              7.7       &  $-$406.626745855377    &               9.7       &  $-$627.480644193817    &                13.0       &  $-$154.076250986245   \\
              7.8       &  $-$493.261808207445    &               9.8       &  $-$607.236006360562    &              13.5       &  $-$123.313365467909   \\
              7.9       &  $-$564.543330776862    &               9.9       &  $-$586.808634364519    &                14.0       &  $-$99.0517088984615   \\
              8.0       &  $-$622.296924164693    &              10.0       &  $-$566.344786996666    &              14.5       &  $-$79.8868022335593   \\
              8.1       &  $-$668.163414255055    &              10.1       &  $-$545.963873169331    &                15.0       &  $-$64.7056634495449   \\
              8.2       &  $-$703.629551622843    &              10.2       &  $-$525.767267986650    &              15.5       &  $-$52.6401140333373   \\
              8.3       &  $-$730.026393137440    &              10.3       &  $-$505.840657675087    &                16.0       &  $-$43.0174850895937   \\
              8.4       &  $-$748.547347148735    &              10.4       &  $-$486.253895514174    &              16.5       &  $-$35.3133121226139   \\
              8.5       &  $-$760.256868094275    &              10.5       &  $-$467.064631313888    &                17.0       &  $-$29.1215244271569   \\
              8.6       &  $-$766.106999612316    &              10.6       &  $-$448.319777407923    &              17.5       &  $-$24.1273429014569   \\
              8.7       &  $-$766.942290212839    &              10.7       &  $-$430.056805726134    &                18.0       &  $-$20.0834953954821   \\
              8.8       &  $-$763.511447018345    &              10.8       &  $-$412.304806381708    &              18.5       &  $-$16.7958394274081   \\
              8.9       &  $-$756.477032518151    &              10.9       &  $-$395.085506635547    &                19.0       &  $-$14.1139773714497   \\
              9.0       &  $-$746.423038708948    &              11.0       &  $-$378.414408279889    &              19.5       &  $-$11.9094506999236   \\
              9.1       &  $-$733.862648819818    &              11.2       &  $-$346.752289713281    &                20.0       &  $-$10.0943440302398 \\
              9.2       &  $-$719.246967350335    &              11.4       &  $-$317.346954936712    & 
\end{tabular}
\end{table*}  

\begin{table*} 
\caption{Potential energy $V_{\rm ss}(R)$ of the energetically-lowest
``spin-stretched'' state of Tm$_2$ as function of
internuclear separation $R$. The $R$-dependent uncertainty of this potential is discussed in the text.}
\label{tab:TmSS}
	\begin{tabular}{c@{~}c@{~~}|@{~}c@{~}c@{~~}|@{~}c@{~}c}
	$R/a_0$  &   $V_{\rm ss}/hc$ (cm$^{-1}$) &
		$R/a_0$  &   $V_{\rm ss}/hc$ (cm$^{-1}$) &
		$R/a_0$  &   $V_{\rm ss}/hc$ (cm$^{-1}$) 
	\\
\hline
6.25 & 3874.15018867483 & 8.6 & $-$775.049924968957 & 10.75 & $-$409.392022314815 \\
6.5 & 2452.88195449737 & 8.7 & $-$772.46879342323 & 11.0 & $-$367.345465164348 \\
6.75 & 1398.48143296268 & 8.75 & $-$769.709228982994 & 11.25 & $-$328.942955195758 \\
6.8 & 1224.59520643307 & 8.8 & $-$766.073434020327 & 11.5 & $-$294.175889173847 \\
6.9 & 909.16853126916 & 8.9 & $-$756.484235631714 & 11.75 & $-$262.878651327859 \\
7.0 & 633.431409261034 & 9.0 & $-$744.222911614132 & 12.0 & $-$234.883475377653 \\
7.1 & 393.453268965899 & 9.1 & $-$729.769650342498 & 12.25 & $-$209.924884847286 \\
7.2 & 185.614885631341 & 9.2 & $-$713.532718742361 & 12.5 & $-$187.720788736913 \\
7.25 & 92.6843395024816 & 9.25 & $-$704.864106741569 & 12.75 & $-$168.003713562084 \\
7.3 & 6.56791035932326 & 9.3 & $-$695.884191964824 & 13.0 & $-$150.499513340797 \\
7.4 & $-$146.706530584091 & 9.4 & $-$677.132015128778 & 13.5 & $-$121.195083284982 \\
7.5 & $-$276.986305870368 & 9.5 & $-$657.551518907663 & 14.0 & $-$98.0218311326522 \\
7.6 & $-$386.841551372091 & 9.6 & $-$637.379384988748 & 14.5 & $-$79.6165119880272 \\
7.7 & $-$478.589634023197 & 9.7 & $-$616.821173105488 & 15.0 & $-$64.9113377098249 \\
7.75 & $-$518.306204431474 & 9.75 & $-$606.45387137768 & 15.5 & $-$53.1237496407613 \\
7.8 & $-$554.273972390536 & 9.8 & $-$596.054548248776 & 16.0 & $-$43.6507851515064 \\
7.9 & $-$615.850590443533 & 9.9 & $-$575.229126068204 & 16.5 & $-$36.0202004857629 \\
8.0 & $-$664.991862988103 & 10.0 & $-$554.467856276277 & 17.0 & $-$29.8546529570767 \\
8.1 & $-$703.262622294269 & 10.1 & $-$533.883746459349 & 17.5 & $-$24.8547571665504 \\
8.2 & $-$732.03585776905 & 10.2 & $-$513.570051395331 & 18.0 & $-$20.7752063343089 \\
8.25 & $-$743.254940662834 & 10.25 & $-$503.537338666485 & 18.5 & $-$17.4646289323737 \\
8.3 & $-$752.551798724083 & 10.3 & $-$493.597463728117 & 19.0 & $-$14.7468524614891 \\
8.4 & $-$765.919099466286 & 10.4 & $-$474.031826052711 & 19.5 & $-$12.5153879327555 \\
8.5 & $-$773.126032186692 & 10.5 & $-$454.926141284319 & 20.0 & $-$10.664119330474 

\end{tabular}
\end{table*}  

The spin-stretched potential for ${R<R_{\rm min}}$ is found by
linear extrapolation using the first two separations larger than or equal to $R_{\rm min}$.  For ${R> R_{\rm disp}}$ with ${R_{\rm
disp}>R_{\rm max}}$ we use the dispersive form 
\begin{equation}
V_{\rm disp}(R)=C_{6,\rm ss}/R^6 +C_{8,\rm ss}/R^8+C_{10,\rm ss}/R^{10}\,,
\end{equation} 
where the van-der-Waals coefficient $C_{6,\rm ss}$ is 
\begin{equation}
	C_{\rm ss}=C_0^{(1)}   +(2j)(2j-1)\, C_2^{(1)}/\sqrt{6}
	\label{eq:css}
\end{equation}
with $j=6$ and 7/2 for Er$_2$ and Tm$_2$, respectively based on two relevant values for $C_k^{(i)}$ are given
in the table in the main text and the derivation in this Appendix \ref{app:nonrel}. 
Coefficients $C_{8,\rm ss}$ and $C_{10,\rm ss}$ are fixed such that the dispersive form agrees with 
the potential energy from coupled-cluster theory at the two largest radial points $R\le R_{\rm max}$. 
We use $R_{\rm disp}=R_{\rm max}+0.5a_0$ for
both Er$_2$ and Tm$_2$, add $(R_{\rm disp}, V_{\rm
disp}(R_{\rm disp}))$ to the coupled-cluster data, and for $R\in(R_{\rm
min},R_{\rm disp})$ interpolate this extended coupled-cluster data
set times $R^6$ using the Akima spline \cite{Akima1991}.  
The function $R^6 V_{\rm ss}(R)$ varies significantly less than $V_{\rm ss}(R)$.
The fitted $C_{8,\rm ss}$ are consistent with typical values  based on the induced 
quadrupole-quadrupole interaction for other di-atomic molecules \cite{Tao2012}
and the contributions from the five omitted dispersion terms is small compared to
the uncertainties in the potentials.

The uncertainty budget of  $V_{\rm ss}(R)$ as function of $R$ has two components.  
The first is the complete basis set error of the RCCSD(T) calculations.  The second is that
four-electron excitations might need to be included  in  our
open-shell molecules. This corresponds to accounting for non-perturbative
triple as well as quadruple excitations.  Basis-set superpositions
errors increase the depth of the potentials $V_{\rm ss}(R)$, while
non-perturbative triple and quadruple corrections often are of
opposite sign, nearly cancel, but lead to shallower potentials for
dimers \cite{Feller2007,Kodrycka2019}. Here, based on
the differences of calculations with triple- and quadruple-zeta accuracy
basis sets, we assume that the one-standard-deviation uncertainties
of  $V_{\rm ss}(R)$ is $2\times u(C_0^{(1)})/R^6$ for $R<R_{\rm
max}$, where $u(C_0^{(1)})$ is the one-standard-deviation uncertainty
of the  isotropic dispersion coefficient $C_0^{(1)}$.
    

\section{ Relativistic configuration-interaction calculations and expansion in spin tensor operators}

We  have used the  direct relativistic configuration interaction (DIRRCI) method in DIRAC 2019 \cite{DIRAC19} to determine the 
energy splittings among the relativistic adiabatic potential curves of Er$_2$ and Tm$_2$  for $R\le 12a_0$. Basis sets have been described in Sec.~\ref{sec:method}. In Sec.~\ref{sec:method}, we also described how the spin-stretched potential  $V_{\rm ss}(R)$ 
and the energy splittings are used to construct relativistic adiabatic potential curves $V_{\rm rel}(R;n,\Omega_{g/u}^\pm)$.
We find a common uncorrelated one-standard-deviation uncertainty $u(R)=hc\times10$ cm$^{-1}$ independent of $R$ for
all  potential energy splittings. This follows from a comparison of  DIRRCI calculations with different basis set size. 
In addition, the uncertainty in the splittings and that of $V_{\rm ss}(R)$ are uncorrelated.

The $V_{\rm rel}(R;n,\Omega_{g/u}^\pm)$ were fit to an expansion in terms of seven spin-spin tensor operators with strengths
$V_k^{(i)}(R)$ defined in the main text. Only $V_0^{(1)}(R)$ and $V_2^{(1)}(R)$ were found to be statistically relevant and we 
finally decided to only present results with those two strengths as fitting parameters with the remaining five strengths set to zero.
The reduced chi-square $\chi^2_\nu$ of this adjustment is less than one for all $R<12a_0$ so that the fit is consistent. 

Tables~\ref{tab:anisoEr2} and \ref{tab:anisoTm2} contain values of the  spin tensor strengths $V_0^{(1)}(R)$ and $V_2^{(1)}(R)$ as functions of interatomic separations for Er$_2$ and Tm$_2$, respectively. Note
that as the uncertainty in the splittings and that of $V_{\rm ss}(R)$ are uncorrelated strictly speaking strength $V_{2}^{(1)}(R)$ is the only adjusted constant and 
\begin{equation}
	V_0^{(1)}(R) =  V_{\rm ss}(R)-(2j)(2j-1) V_2^{(1)}(R)/\sqrt{6}\,,
	\label{eq:v01}
\end{equation}
where $j=6$ and 7/2 for Er$_2$ and Tm$_2$, respectively.
The spin-tensor strength $V_{2}^{(1)}(R)$ has an one-standard-deviation uncertainty of  $hc\times 0.094$ cm$^{-1}$ and
 $hc\times 0.40$ cm$^{-1}$ for Er$_2$ and Tm$_2$  independent of $R$, respectively. 
 The uncertainty of $V_0^{(1)}(R) $ follows from error
propagation of Eq.~(\ref{eq:v01}).  The contribution from $V_{\rm
ss}(R)$ always dominates. In addition, the absolute value of the
difference between $V_0^{(1)}(R)$ and $V_{\rm ss}(R)$ are no larger
than the uncertainty of $V_{\rm ss}(R)$ for all $R$. Hence, we
surmise that the relativistic corrections to the non-relativistic
spin-stretched potential $V_{\rm ss}(R)$ are of similar magnitude
as well.

For ${R>12a_0}$ the relativistic configuration-interaction calculations do not
converge.  As shown in Fig.~4 in the main text,  however, the adjusted
anisotropic strengths $V_{2}^{(1)}(R)$ at ${R=12a_0}$ for the two dimers are
already consistent, {\it i.e.} within our uncertainties, with its
asymptotic  van-der-Waals $C_{2}^{(1)}/R^6$ behavior.  We then use the van-der-Waals behavior
for $R>R_{\rm rel}$ with $R_{\rm rel}=12a_0+0.5a_0$. A smooth
connection of the strength between $12a_0$ and  $R_{\rm rel}$ is
ensured by adding point $(R_{\rm rel},C_{2}^{(1)}/R_{\rm rel}^6)$
to the $R\le 12a_0$ fitted values for  $V_{2}^{(1)}(R)$ and
interpolate $R^6 \,V_{2}^{(1)}(R)$ with the Akima spline \cite{Akima1991}.

Finally, even though the five weaker $V_k^{(i)}(R)$ are consistent with
zero in the least-squares adjustment, we use $V_k^{(i)}(R)=C_k^{(i)}/R^6$ 
for all $R$ for these five strengths in the calculation of the rovibrational levels
of Er$_2$ and Tm$_2$.


\begin{table} [h]
	\caption{Spin tensor strengths $V_0^{(1)}(R)$ and $V_2^{(1)}(R)$ for Er$_2$ 
	as functions of separation $R$ for $R\le12a_0$.  The isotropic strength $V_0^{(1)}(R)$
	is found from Eq.~(\ref{eq:v01}). Its $R$-dependent uncertainty equals that of the potential of the spin-stretched state.
	The uncertainty
	of  $V_2^{(1)}(R)$ is $hc\times 0.094$ cm$^{-1}$  independent of $R$.}
	\label{tab:anisoEr2}
	\begin{tabular}
	{r@{\quad}l@{\quad}|@{\quad}l}
		$R/a_0$ &  $V_0^{(1)}/hc$  & $V_2^{(1)}/hc$  \\
		   & (cm$^{-1}$) &  (cm$^{-1}$) \\
		\hline
		7.2 &  & 0.75334948\\
 7.4 & $-$86.14282611 &   0.97861800 \\ 
 7.6 & $-$362.33159884 &   1.10797570 \\ 
 7.8 & $-$555.79268757 &   1.16036930 \\ 
 8.0 & $-$684.70138046 &   1.15802330 \\ 
 8.2 & $-$763.82004690 &   1.11693940 \\ 
 8.4 & $-$805.15566699 &   1.05046590 \\ 
 8.6 & $-$818.30347406 &   0.96859643 \\ 
 8.8 & $-$810.88531912 &   0.87910465 \\ 
 9.0 & $-$788.86447184 &   0.78757466 \\ 
 9.2 & $-$756.85758280 &   0.69793043 \\ 
 9.4 & $-$718.37785998 &   0.61237588 \\ 
 9.6 & $-$676.09538548 &   0.53262488 \\ 
 9.8 & $-$632.01638040 &   0.45984297 \\ 
10.0 & $-$587.59296366 &   0.39429690 \\ 
10.2 & $-$543.84435218 &   0.33545176 \\ 
10.4 & $-$501.52123195 &   0.28331200 \\ 
10.6 & $-$461.13522634 &   0.23781296 \\ 
10.8 & $-$423.00143547 &   0.19849457 \\ 
11.0 & $-$387.29369989 &   0.16477071 \\ 
11.2 & $-$354.08026901 &   0.13598341 \\ 
11.4 & $-$323.36078965 &   0.11159717 \\ 
11.6 & $-$295.07053724 &   0.09093034 \\ 
11.8 & $-$269.11169606 &   0.07346497  \\
 12.0  &    & 0.05871469
\end{tabular} 
\end{table}  

\begin{table} [h]
	\caption{Spin tensor strengths $V_0^{(1)}(R)$ and $V_2^{(1)}(R)$ for 
	Tm$_2$ as functions of separation $R$ for $R\le12a_0$. The isotropic strength $V_0^{(1)}(R)$
	is found from Eq.~(\ref{eq:v01}). Its $R$-dependent uncertainty equals that of the potential of the spin-stretched state. The uncertainty
	of  $V_2^{(1)}(R)$ is $hc\times 0.40$ cm$^{-1}$ independent of $R$.}
	\label{tab:anisoTm2}
	\begin{tabular}
	{r@{\quad}l@{\quad}|@{\quad}l}
		$R/a_0$ &  $V_0^{(1)}/hc$  & $V_2^{(1)}/hc$  \\
		   & (cm$^{-1}$) &  (cm$^{-1}$) \\
		\hline
 7.0 & 594.90655204 &   2.2468153 \\ 
 7.2 & 133.54032587 &   3.0370500 \\ 
 7.4 & $-$206.86711784 &   3.5086367 \\ 
 7.6 & $-$450.80090236 &   3.7301851 \\ 
 7.8 & $-$618.71806662 &   3.7584559 \\ 
 8.0 & $-$727.59730629 &   3.6512236 \\ 
 8.1 & $-$764.32796312 &   3.5614030 \\ 
 8.2 & $-$791.25843938 &   3.4539311 \\ 
 8.3 & $-$809.69858744 &   3.3328684 \\ 
 8.4 & $-$820.81468219 &   3.2015754 \\ 
 8.5 & $-$825.64585040 &   3.0630180 \\ 
 8.6 & $-$825.11351391 &   2.9197678 \\ 
 8.7 & $-$820.03654142 &   2.7742074 \\ 
 8.8 & $-$811.13357004 &   2.6279605 \\ 
 8.9 & $-$799.04967438 &   2.4824668 \\ 
 9.0 & $-$784.32927135 &   2.3390504 \\ 
 9.1 & $-$767.46932713 &   2.1986898 \\ 
 9.2 & $-$748.89238132 &   2.0622174 \\ 
 9.3 & $-$728.98072835 &   1.9302292 \\ 
 9.4 & $-$708.05036358 &   1.8031947 \\ 
 9.5 & $-$686.38223486 &   1.6814415 \\ 
 9.6 & $-$664.21695040 &   1.5651986 \\ 
 9.8 & $-$619.19702752 &   1.3496968 \\ 
10.0 & $-$574.30698185 &   1.1570413 \\ 
10.2 & $-$530.48373364 &   0.98642598 \\ 
10.4 & $-$488.37811753 &   0.83669271 \\ 
11.0 & $-$375.85614024 &   0.49635265 \\ 
12.0 & $-$238.11360714 &   0.18838511 
\end{tabular} 
\end{table}  

\begin{figure} [h]
    \centering
    \includegraphics[scale=0.175,trim= 5 0 0 40,clip]{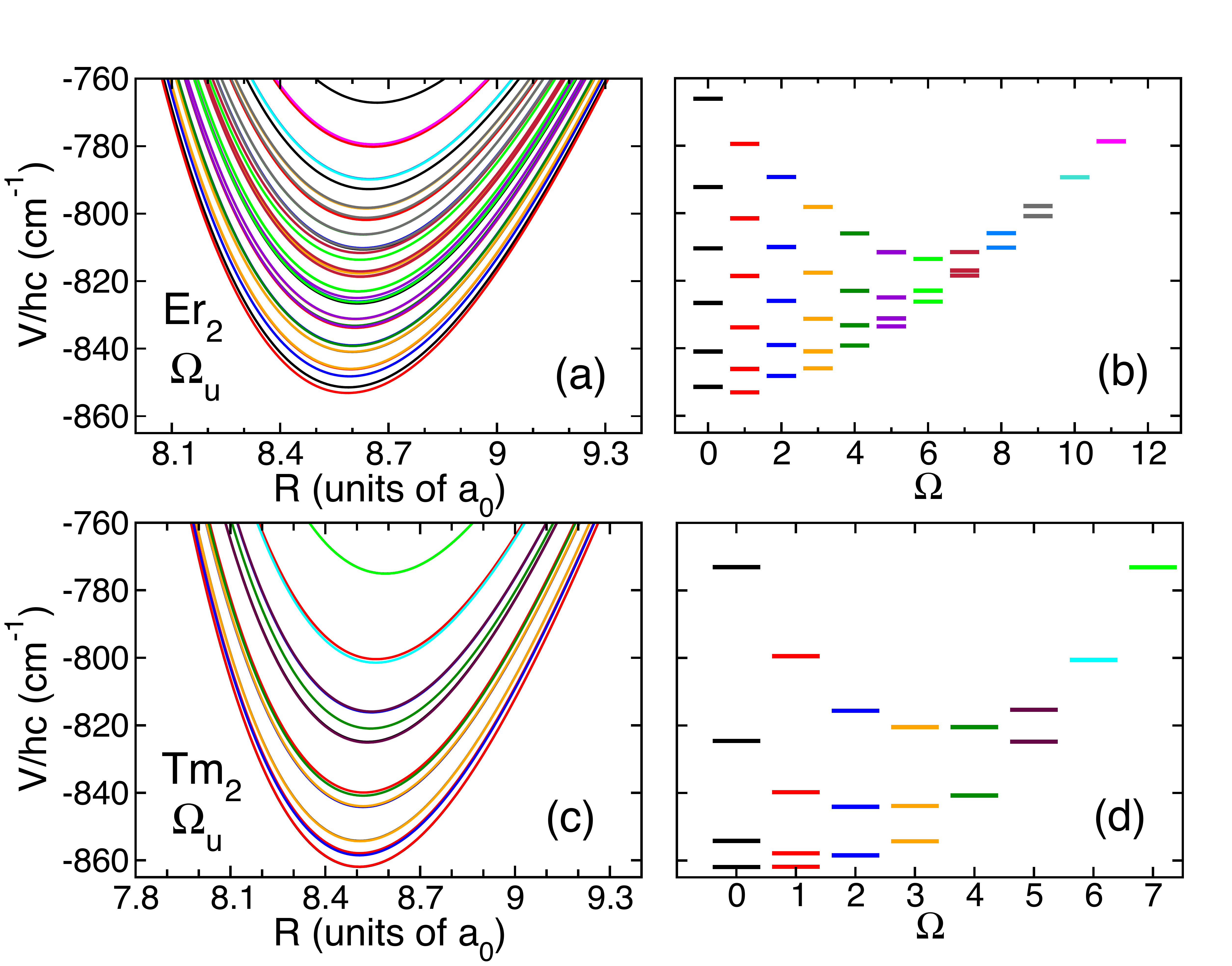}
    \caption{Relativistic $\Omega_u^{\pm}$ potential energy
curves with {\it ungerade} symmetry for Er$_2$ (panel a) and Tm$_2$
(panel c) as functions of internuclear separation $R$ near the
equilibrium separation as obtained from electronic structure
calculations. All potentials approach zero energy for $R\to\infty$.
Panels (b) and (d) show potential energies from panels (a) and (c)
using the same line colors at  the equilibrium separation as
functions of projection quantum number $\Omega$ for Er$_2$ and
Tm$_2$, respectively.  {\it Ungerade} $\Omega=0$ states are $0^-$
states.  }
\label{fig:Er2_Tm2_ungerade_omega}
\end{figure}

Figure \ref{fig:Er2_Tm2_ungerade_omega} shows the {\it ungerade} potentials of
Er$_2$ and Tm$_2$ near their equilibrium separation. The figure  complements the figure with {\it gerade} state potentials in the main text 

\section{Derivation of the dispersion potentials}

Long-range van-der-Waals dispersion interactions have been derived and studied in many 
settings \cite{Stone,Derevianko2010,Kotochigova2011}. We repeat part of these derivations in order to explain the 
relationships among the strengths of the seven spin-tensor operators contributing 
the atom-atom interaction $V(\vec {\mathbf R})$ for large separations $R$.
In this section we rely on Ref.~\cite{Brink1993} for notation regarding angular 
momentum operators as well as the manipulation of these operators 
with one clearly-stated exception regarding reduced matrix elements.

We consider  interacting atoms with electronic eigenstates $| n\, b
\beta \rangle$ with total electronic angular momentum quantum numbers  $b$, projection quantum numbers $\beta$ on a space- or laboratory-fixed axis, and energies $E_{nb}$ that 
are independent of $\beta$. Label $n$ further uniquely specifies states. 
The ground state of the atoms is $| g\,j m\rangle$.  

The van-der-Waals potential operator between two ground-state
atoms, $| g_1\,j_1 m_1,\,g_2\,j_2m_2\rangle=| g_1\,j_1 m_1\rangle|g_2\,j_2m_2\rangle$, is derived from (degenerate) second-order perturbation theory in
the anisotropic electric dipole-dipole interaction $V_{\rm dd} ({\bf \vec R})$ and is given by
\begin{eqnarray}
 \lefteqn{ V_{\rm vdW}({\bf \vec R})=  \sum'_{n_1 b_1 \beta_1,\ n_2 b_2 \beta_2}
     V_{\rm dd}({\bf \vec R}) | n_1 b_1 \beta_1,n_2 b_2 \beta_2\rangle
         } 
         \label{eq:dipdip}\\
         && 
         \times
        \frac{1}{E_{g_1j_1}+E_{g_2j_2}-E_{n_1b_1}-E_{n_2b_2}}
        \langle n_1 b_1 \beta_1,n_2 b_2 \beta_2| V_{\rm dd}({\bf \vec R}) \,,
        \nonumber
\end{eqnarray}
where
\begin{eqnarray}
    V_{\rm dd} ({\bf \vec R})&=&  -\frac{1}{4\pi\epsilon_0} \sqrt{6} \frac{1}{R^3}  T_2(d_1, d_2)
     \cdot C_2({\bf \hat R})\,,
\end{eqnarray}
$ d_i$ is the rank-1 electric dipole moment operator of atom $i=1$ or 2,  $C_{kq}({\bf \hat R})$ is a spherical harmonic, and rank-$k$ spherical-tensor operator $T_{kq}(R, S)\equiv[ R \otimes S ]_{kq}$
with the $\otimes$ notation of \cite{Santra2003} is constructed from spherical-tensor operators $R$ and $S$ with rank $r$ and $s$, respectively.
The prime on the sums in Eq.~(\ref{eq:dipdip}) indicates that the
sums exclude the term where both atoms are in the ground state
and $b_i=|j_i-1|,\dots,j_i+1$.  
The energy denominator  is negative and does not depend on  projection
quantum numbers.
Finally, $\epsilon_0$ is the vacuum electric permittivity.

The sums in Eq.~(\ref{eq:dipdip}) can be rearranged in several steps
using Appendix VI of Ref.~\cite{Brink1993}. As operators $d_1$ and $d_2$ commute, we first note that
\begin{eqnarray}
\lefteqn{
 (T_2(d_1,d_2)\cdot C_2)(T_2(d_1,d_2)\cdot C_2)=}\\
&&
\quad   \quad  
 \sum_{k} (-1)^k\, T_k(\,T_2(d_1,d_2),T_2(d_1,d_2)\,) \cdot T_k(C_2,C_2)
\nonumber
      \label{eq:step1}
\end{eqnarray}
with ${T_{kq}( C_2,C_2)= \langle k 0 | 22 00 \rangle C_{kq}}$ and 
Clebsch-Gordan coefficient $\langle j_3m_3 | j_1 j_2 m_1m_2\rangle$. We omitted  the dependence on orientation ${\bf \hat R}$ of the spherical harmonics for clarity.
The right hand side of Eq.~(\ref{eq:step1}) is only nonzero  for even $k=0$, 2, or 4. Secondly, we note that
\begin{eqnarray}
 \lefteqn{
 T_{kq}(\,  T_2(d_1,d_2), T_2(d_1,d_2)\,)= 5 \sum_{l_1l_2} \sqrt{(2l_1+1)(2l_2+1)} 
}
\nonumber \\
&& \quad\quad 
  \times
     \left\{
         \begin{array}{ccc}
              1 & 1 & l_1\\  1 & 1 & l_2 \\ 2 & 2 & k
         \end{array}
      \right\} T_{kq}( \,T_{l_1}(d_1,d_1), T_{l_2}(d_2,d_2)\,) \,,
      \quad \quad\quad
      \label{eq:step2}
\end{eqnarray}
where the $d_i$ have been grouped by atom
and $l_i=0,1$, or 2 and $l_1+l_2$ is even for a nonzero value
of the nine-$j$ symbol 
$\renewcommand{\arraystretch}{0.5}\left\{\!\!\begin{array}{c}\cdots\\\cdots\\\cdots\end{array}\!\!\right\}$.

Next, we isolate the sums over projection quantum numbers and labels $n$ in Eq.~(\ref{eq:dipdip}).  For  atom $i$, we  define spherical tensor operators
\begin{eqnarray}
 B_{lq}(b,j;i) &=  & \sum_{\beta} 
  T_{lq}\left(
   \frac{d_i|nb\beta\rangle }{\langle gj||d_i || n b\rangle},
   \frac{\langle nb\beta|d_i}{\langle nb || d_i || gj\rangle} \right)\quad
\end{eqnarray}
with rank ${l=0}$, 1, or 2 and ${b=|j-1|,\dots,j+1}$.
Here, $\langle gj||d_i || n b\rangle$ and $\langle nb || d_i || gj\rangle$ are reduced matrix elements of the electric dipole moment operator between the atomic ground state and 
 excited state $| n\, b \beta \rangle$.
Crucially, for a ground-state atom  we derive that 
\begin{eqnarray}
\lefteqn{
  \langle g j m| B_{lq}(b,j;i)| g j m'\rangle= }       \label{eq:Atensor}
\\
                 &&\qquad \langle  jm | j l m' q\rangle \sqrt{(2b+1)(2l+1)} W( j 1 j 1; b l)
               \nonumber 
            \end{eqnarray}
based on the Wigner-Eckart theorem,  Eq.~(3.12) of Ref.~\cite{Brink1993}, and 
symmetries of the Racah symbol $W( abcd; ef)$. These matrix elements
are independent of label $n$ of the excited state, but still depend on  its total angular momentum $b$.

Moreover, using the Wigner-Eckart theorem again,
we realize that the $m$, $m'$, and $q$ dependences of \\ $\langle g j m| B_{lq}(b,j;i)| g j m'\rangle$ 
are identical to those for the identity operator, atomic angular momentum operator
$j_q$, and dipole operator $T_{2q}(j,j)$ for $l=0$, 1, and 2, respectively.
In fact, we find 
\begin{eqnarray}
   B_{lq}(b,j;i) &=& O_{lq}(i)  M(b,j)
\end{eqnarray}
with rank-$l$ operator $O_{lq}(i) = I$, $j_{iq}/\hbar$, and $T_{2q}(j_i,j_i)/\hbar^2$
for atom $i$ and $l=0$, 1, and 2, respectively.  
Here, $\hbar$ is the reduced Planck constant and
the function $M(b,j;l)$ is given by
 \begin{eqnarray}
  M(b,j;l=0) &=&(-1)^{b-j+1} \frac{1}{\sqrt{3}}\sqrt{\frac{2b+1}{2j+1}}\,,
\end{eqnarray}
\begin{equation}
   M(b,j;l=1) = 
      \frac{(-1)^{b-j} }{2\sqrt{2}}\sqrt{\frac{2b+1}{2j+1}} \frac{2+j(j+1)-b(b+1)}{j(j+1)}
     \,,
\end{equation}
and
\begin{eqnarray}
   M(b,j;l=2) &=& \sqrt{\frac{2b+1  }{2j+1}}
      \frac{ W( j 1 j 1; b 2) }{W( j 1 j 1; j 2)}\frac{1}{j(j+1)}
      \,.\quad
\end{eqnarray}

We  put everything together to find for two ground-state atoms
\begin{eqnarray}
 \lefteqn{       V_{\rm vdW}({\bf \vec R}) =\frac{1}{R^6}       
   \sum_{k=0,2,4}    \sum_{l_1l_2} (T_{k}\left( O_{l_1}(1), O_{l_2}(2)\right)\cdot C_k({\bf \hat R}) )}\nonumber \\
  && \times 30 \,
  \sqrt{\frac{(2l_1+1)(2l_2+1) }{(2j_1+1)(2j_2+1)}} 
     \left\{
         \begin{array}{ccc}
              1 & 1 & l_1\\  1 & 1 & l_2 \\ 2 & 2 & k
         \end{array}
      \right\}  \langle k0|2200\rangle \nonumber
         \\
      && \quad  \quad\quad \times 
      \sum_{b_1=|j_1-1|}^{j_1+1}
      \sum_{b_2=|j_2-1|}^{j_2+1}   
        \frac{(-1)^{b_1-j_1} (-1)^{b_2-j_2} }{\sqrt{(2b_1+1)(2b_2+1)}}
                      \label{eq:finalvdW}\
          \\
 \  &&   \quad\quad \quad\quad \quad\quad\quad\quad \times 
    \, M(b_1,j_1;l_1) M(b_2,j_2;l_2)   \, h_{b_1b_2}
      \,, \nonumber
\end{eqnarray}
where the matrix
\begin{equation}
        h_{b_1b_2}=  \frac{1}{(4\pi\epsilon_0)^2} \sum'_{n_1n_2}
  \frac{ (g_1j_1||d_1 || n_1 b_1)^2
       \,  (g_2j_2||d_2 || n_2 b_2)^2}{E_{g_1j_1}+E_{g_2j_2}-E_{n_1b_1}-E_{n_2b_2}}
       \label{eq:h12}
\end{equation}
or, equivalently,
\[
          E_{\rm h} a_0^6 \sum'_{n_1n_2}
  \frac{ (g_1j_1||d_1/(ea_0) || n_1 b_1)^2
       \,  (g_2j_2||d_2/(ea_0) || n_2 b_2)^2}{(E_{g_1j_1}+E_{g_2j_2}-E_{n_1b_1}-E_{n_2b_2})/E_{\rm h}}
\]
is symmetric for homonuclear dimers. Here,  $e$ is the elementary charge, $E_{\rm h}$ is the Hartree energy,  and
$a_0$ is the Bohr radius. The prime in the sums over labels $n_1$ and $n_2$ excludes the case where
both atoms are in their ground state and we have introduced the more symmetric
reduced matrix elements $( j|| d || j')=(-1)^{j-j'} (j'||d||j)^*=\sqrt{2j+1}\langle j || d || j'\rangle$ used by, for example, Edmonds in Ref.~\cite{Edmonds1957}.

The allowed values for $k$, $l_1$, and $l_2$ and the operators on the 
first line of Eq.~(\ref{eq:finalvdW}) lead to the seven spin-tensor
operators defined  in the main text.
The  last three lines of Eq.~(\ref{eq:finalvdW}) correspond to the van-der-Waals coefficients $C_k^{(i)}$.
For example, the choice $O_{l_1}(1)=j_{1}/\hbar$ and $O_{l_2}(2)=j_{2}/\hbar$  leads to spin-tensor operators
$[j_1 \otimes j_2 ]_{k0}/\hbar^2$ with $k=0$ or 2 in the main text using the $\otimes$ notation of
Ref.~\cite{Santra2003} for combining spherical tensor operators. This notation is equivalent
to the notation used in Ref.~\cite{Brink1993}.
Further analysis of Eq.~(\ref{eq:finalvdW}) shows that the operators
with the same $l_1$ and $l_2$ but different $k$ have related van-der-Waals coefficients as the $k$ dependence is isolated in the nine-$j$ symbol
and the Clebsch-Gordan coefficient in the second  line. This leads to the relationships 
between the two spin-tensors with $l_1=l_2=1$ and the three spin-tensors with $l_1=l_2=2$
given in the main text.

Lists of currently available atomic transition energies $E_{nb}-E_{gj}$ and observed Einstein $A$ coefficients or oscillator strengths $f$
for erbium and thulium atoms are given in Tables \ref{tab:ErA}, \ref{tab:TmOne}, and \ref{tab:TmTwo} below.
The relevant relationships between the reduced matrix elements $(gj||d || n b)$ in Eq.~(\ref{eq:h12}) on the one hand and  $A$ and $f$ on the other are
\begin{equation}
        A_{n b\to g j}
               =
          \frac{4}{3} \frac{E_{\rm h}}{\hbar} \alpha^3
          \left(\frac{{E_{nb}-E_{g j}}}{E_{\rm h}}\right)^3
          \left|( g\, j|| \frac{d}{ea_0} || n\,b)\right|^2
          \frac{1}{2b+1}
\end{equation}
and
\begin{eqnarray}
        f_{gj,nb}
        &     = &
                \frac{2}{3} \frac{E_{nb}-E_{g j}}{E_{\rm h}}
                \left|( g\, j || \frac{d}{ea_0} || n\,b )\right|^2
                \frac{1}{2j+1}\,,\quad
\end{eqnarray}
respectively. Here, $\alpha$ is the fine-structure constant.

For Er$_2$ atomic transition data or lines from 48 excited states
to the ground state are available. The majority of the data is taken
from Refs.~\cite{Lawler2010,Meggers1975}. References \cite{Gorshkov1981}
and \cite{Komarovskii1993} each supply one line, while for two other
lines we rely on private communications \cite{Lawler2018,Ferlaino2015}.
For Tm$_2$ atomic transition data  from 65 excited states
to the ground state are available.  Data have been taken
from Refs.~\cite{Wickliffe1997,Kramida2019,Penkin1976}.
When line strength information of a transition is available from more than one source,
the most accurate datum is chosen.

The uncertainties of and correlations among the van-der-Waals coefficients follow from 
error propagation on Eqs.~(\ref{eq:finalvdW}) and (\ref{eq:h12}) and are 
dominated by the uncorrelated uncertainties of the Einstein $A$ coefficients and oscillator strengths.
Uncertainties in the transition energies give negligible contributions.
Our values for the van-der-Waals coefficients are listed in the Table in the main text.
Their covariances can be found in Table~\ref{tab:cc}  below.

\begin{table*}[h]
	\caption{Atomic transition energies and Einstein $A$
	coefficients from exited states of erbium to its $j^\prime=6$
	ground state.  
	Columns labeled $\Delta E$, $A$, $u(A)$, and
	$j$ give transition energies, the value and one-standard
	deviation uncertainty of the Einstein $A$ coefficients, and
	the total electronic angular momenta $j$ of the excited
	states, respectively. A reference to the original data is
	given in columns labeled by ``Ref.''. }
\label{tab:ErA}
	\begin{tabular}{c@{~~}|@{~~}rr@{~~}|@{~~}c@{~~}|@{~}c||@{~~}c@{~~}|@{~~}rr@{~~}|@{~~}c@{~~}|@{~}c}
	$\Delta E/hc$  &  \multicolumn{1}{c}{$A$}  & \multicolumn{1}{c|@{~~}}{$u(A)$}  & $j$ & Ref. &
	$\Delta E/hc$  &  \multicolumn{1}{c}{$A$}  & \multicolumn{1}{c|@{~~}}{$u(A)$}  & $j$ & Ref. \\
	(cm$^{-1}$)  & \multicolumn{2}{c|@{~~}}{($10^6$ s$^{-1}$)} & & &
    (cm$^{-1}$)  & \multicolumn{2}{c|@{~~}}{($10^6$ s$^{-1}$)} & \\ 
	\hline
	11401.197  & 0.006377  & 0.00159425 &  5  &   \cite{Meggers1975}  &  	23885.406  &     1.02  &       0.06  &  5  &   \cite{Lawler2010} \\
	11799.778  &  0.01076  &    0.00269  &  6  &   \cite{Meggers1975}  &  	24083.166  &     102.  &         5.  &  5  &   \cite{Lawler2010} \\
	11887.503  &  0.01539  &  0.0038475  &  7  &   \cite{Meggers1975}  &  	24457.139  &     32.6  &        1.6  &  6  &   \cite{Lawler2010} \\
	15185.352  &   0.1431  &   0.035775  &  5  &   \cite{Meggers1975}  &  	24943.272  &     220.  &        10.  &  7  &   \cite{Ferlaino2015}\\
	15846.549  &   0.2624  &     0.0656  &  7  &   \cite{Meggers1975}  &  	25159.143  &     40.3  &        2.1  &  7  &   \cite{Lawler2010} \\
	16070.095  &     0.92  &       0.05  &  6  &   \cite{Lawler2010}  &  	25162.553  &     37.6  &        1.9  &  5  &   \cite{Lawler2010} \\
	16321.110  &  0.09051  &  0.0226275  &  6  &   \cite{Meggers1975}  &  	25268.259  &     3.59  &       0.18  &  6  &   \cite{Lawler2010} \\
	17073.800  &     0.24  &       0.06  &  6  &   \cite{Komarovskii1993}  &  	25392.779  &     31.9  &        1.6  &  6  &   \cite{Lawler2010} \\
	17157.307  &     1.17  &       0.06  &  7  &   \cite{Lawler2010}  &  	25598.286  &     15.1  &        0.8  &  7  &   \cite{Lawler2010} \\
	17347.860  &     0.84  &       0.04  &  5  &   \cite{Lawler2010}  &  	25681.933  &      63.  &         3.  &  5  &   \cite{Lawler2010} \\
	17456.383  &   0.1833  &   0.045825  &  6  &   \cite{Meggers1975}  &  	25880.274  &     122.  &         6.  &  6  &   \cite{Lawler2010} \\
	19201.343  &     0.53  &      0.053  &  5  &    \cite{Lawler2018}  &  	26237.004  &     29.0  &        1.4  &  6  &   \cite{Lawler2010} \\
	19326.598  &    0.663  &    0.16575  &  6  &   \cite{Meggers1975}  &  	28026.045  &     0.59  &       0.05  &  5  &   \cite{Lawler2010} \\
	19508.432  &   0.6392  &     0.1598  &  6  &   \cite{Meggers1975}  &  	28053.943  &     4.33  &       0.22  &  6  &   \cite{Lawler2010} \\
	21168.430  &     1.16  &       0.06  &  7  &   \cite{Lawler2010}  &  	29550.807  &    0.064  &      0.007  &  5  &   \cite{Lawler2010} \\
	21392.817  &     1.26  &       0.06  &  5  &   \cite{Lawler2010}  &  	29794.862  &    0.296  &      0.025  &  5  &   \cite{Lawler2010} \\
	21701.885  &      7.1  &        0.4  &  6  &   \cite{Lawler2010}  &  	29894.203  &     4.10  &       0.29  &  5  &   \cite{Lawler2010} \\
	22124.268  &    0.264  &      0.019  &  5  &   \cite{Lawler2010}  &  	30007.369  &     7.7  &      1.93  &    6  &   \cite{Gorshkov1981} \\
	22583.504  &     2.55  &       0.13  &  6  &   \cite{Lawler2010}  &  	30251.891  &     1.05  &       0.08  &  5  &   \cite{Lawler2010} \\
	22672.766  &     5.52  &       0.28  &  5  &   \cite{Lawler2010}  &  	30380.282  &      4.3  &        0.3  &  5  &   \cite{Lawler2010} \\
	23080.952  &   0.7405  &   0.185125  &  7  &   \cite{Meggers1975}  &  	30600.160  &    0.168  &      0.017  &  5  &   \cite{Lawler2010} \\
	23311.577  &   0.4924  &     0.1231  &  6  &   \cite{Meggers1975}  &  	31442.927  &    0.084  &      0.009  &  5  &   \cite{Lawler2010} \\
	23447.079  &   0.6011  &   0.150275  &  5  &   \cite{Meggers1975}  &  	32062.166  &    0.175  &      0.027  &  5  &   \cite{Lawler2010} \\
	23855.654  &      6.6  &        0.3  &  5  &   \cite{Lawler2010}  &  	33485.216  &      9.6  &        0.7  &  5  &   \cite{Lawler2010} 
\end{tabular}
\end{table*}

\begin{table*}[h]
	\caption{Some thulium excited atomic eigen energies with
	respect to its $j^\prime=7/2$ ground state and oscillator
	strengths $f$ from the ground state to these excited states.
	The first column gives the transition energy.  The second
	and third column are the value and one-standard deviation
	uncertainty of the oscillator strength, respectively. The
	fourth column gives the total electronic angular momentum
	$j$ of the excited state. A reference to the original data
	is given in the last column. Relevant thulium lines for which Einstein
	$A$ coefficients are available can be found in Table \ref{tab:TmTwo}.} \label{tab:TmOne}
	\begin{tabular}{c@{~~}|@{~~}r@{\quad}r@{~~}|@{~~}c@{~~}|@{~}c}
	$\Delta E/hc$  &  \multicolumn{1}{c}{$f$}  &
	\multicolumn{1}{c|@{~~}}{$u(f)$}  & $j$ & Ref. \\
	    (cm$^{-1}$)  &   &  & \\
	\hline 38342.570 &  0.00169 &   0.00338 &   7/2 &
	\cite{Penkin1976} \\ 39019.090 &  0.00098 &   0.00196 &
	9/2 &    \cite{Penkin1976} \\ 39259.920 &  0.00262  &
	0.00525  &  5/2 &    \cite{Penkin1976} \\ 39580.720 &
	0.00822  &  0.01644  &  7/2 &    \cite{Penkin1976} \\
	39847.040 &  0.00192  &  0.00384  &  7/2 &    \cite{Penkin1976}
	\\ 40101.720 &  0.00121  &  0.00242  &  9/2 &    \cite{Penkin1976}
\end{tabular} 
\end{table*}

\begin{table*}[h]
	\caption{Atomic transition energies and Einstein $A$
	coefficients from exited states of thulium to its $j^\prime=7/2$
	ground state. 
	Columns labeled $\Delta E$, $A$, $u(A)$, and
	$j$ give transition energies, the value and one-standard
	deviation uncertainty of the Einstein $A$ coefficients, and
	the total electronic angular momenta $j$ of the excited
	states, respectively. A reference to the original data is
	given in columns labeled by ``Ref.''. 
	Relevant thulium lines for which oscillator strengths
        are available can be found in Table \ref{tab:TmOne}.
	}
	\label{tab:TmTwo}
	\begin{tabular}{c@{~~}|@{~~}rr@{~~}|@{~~}c@{~~}|@{~}c||@{~~}c@{~~}|@{~~}rr@{~~}|@{~~}c@{~~}|@{~}c}
	$\Delta E/hc$  &  \multicolumn{1}{c}{$A$}  & \multicolumn{1}{c|@{~~}}{$u(A)$}  & $j$ & Ref. &
	$\Delta E/hc$  &  \multicolumn{1}{c}{$A$}  & \multicolumn{1}{c|@{~~}}{$u(A)$}  & $j$ & Ref. \\
		(cm$^{-1}$)  & \multicolumn{2}{c|@{~~}}{($10^6$ s$^{-1}$)} & & &
    (cm$^{-1}$)  &  \multicolumn{2}{c|@{~~}}{($10^6$ s$^{-1}$)} & \\ 
	\hline
	16742.237 &  0.147  &   0.02646  &     7/2 &      \cite{Kramida2019}  &  	29260.590 &   5.28  &   0.264    &     7/2 &     \cite{Wickliffe1997} \\
	16957.006 &  0.651  &   0.03255  &     7/2 &     \cite{Wickliffe1997}  &  	29316.690 &   9.80  &   0.49     &     9/2 &     \cite{Wickliffe1997} \\
	17343.374 &  0.388  &   0.03104  &     7/2 &     \cite{Wickliffe1997}  &  	30082.180 &  0.089  &   0.01157  &     5/2 &     \cite{Wickliffe1997} \\
	17613.659 &   1.30  &   0.065    &     9/2 &     \cite{Wickliffe1997}  &  	30124.020 &  0.615  &   0.05535  &     7/2 &     \cite{Wickliffe1997} \\
	17752.634 &   1.09  &   0.0545   &     5/2 &     \cite{Wickliffe1997}  &  	30302.420 &   1.61  &   0.1127   &     5/2 &     \cite{Wickliffe1997} \\
	18837.385 &   2.17  &   0.1085   &     9/2 &     \cite{Wickliffe1997}  &  	30915.020 &   4.29  &   0.2145   &     9/2 &     \cite{Wickliffe1997} \\
	19548.834 &  0.241  &   0.03615  &     5/2 &     \cite{Wickliffe1997}  &  	31431.880 &   3.82  &   0.191    &     5/2 &     \cite{Wickliffe1997} \\
	19748.543 &  0.049  &   0.00882  &     9/2 &      \cite{Kramida2019}  &  	31440.540 &   1.04  &   0.052    &     9/2 &     \cite{Wickliffe1997} \\
	19753.830 &  0.398  &   0.03184  &     7/2 &     \cite{Wickliffe1997}  &  	31510.240 &   15.9  &   1.272    &     7/2 &     \cite{Wickliffe1997} \\
	21120.836 &   2.0  &   0.1      &     7/2 &     \cite{Wickliffe1997}  &  	32174.490 &   1.50  &   0.135    &     5/2 &     \cite{Wickliffe1997} \\
	21161.401 &  0.421  &   0.02105  &     5/2 &     \cite{Wickliffe1997}  &  	32446.260 &   17.5  &   1.225    &     7/2 &     \cite{Wickliffe1997} \\
	21737.685 &  0.518  &   0.0259   &     9/2 &     \cite{Wickliffe1997}  &  	32811.020 &   16.1  &    1.127  &      7/2 &      \cite{Kramida2019} \\
	22791.176 &   3.71  &   0.1855   &     7/2 &     \cite{Wickliffe1997}  &  	33623.780 &   21.7  &   1.085    &     7/2 &     \cite{Wickliffe1997} \\
	22929.717 &   12.0  &   0.6      &     5/2 &     \cite{Wickliffe1997}  &  	34085.200 &   11.3  &   1.13     &     5/2 &     \cite{Wickliffe1997} \\
	23781.698 &   24.3  &   1.215    &     9/2 &     \cite{Wickliffe1997}  &  	34297.170 &   9.44  &   1.1328   &     7/2 &     \cite{Wickliffe1997} \\
	23873.207 &   64.0  &   3.2      &     7/2 &     \cite{Wickliffe1997}  &  	35026.220 &   24.2  &   1.936    &     5/2 &     \cite{Wickliffe1997} \\
	24348.692 &   63.6  &   3.18     &     9/2 &     \cite{Wickliffe1997}  &  	35261.762 & 1.13  &    0.2034  &       5/2 &      \cite{Kramida2019} \\
	24418.018 &   97.9  &   4.895    &     5/2 &     \cite{Wickliffe1997}  &  	37576.866 & 0.46  &    0.0828  &       9/2 &      \cite{Kramida2019} \\
	25656.019 &   2.95  &   0.1475   &     5/2 &     \cite{Wickliffe1997}  &  	37711.074 & 0.38  &    0.0684  &       9/2 &      \cite{Kramida2019} \\
	25717.197 &   37.2  &   1.86     &     7/2 &     \cite{Wickliffe1997}  &  	38120.710 & 5.2  &     0.936  &       9/2 &      \cite{Kramida2019} \\
	25745.117 &    106  &   5.3      &     5/2 &     \cite{Wickliffe1997}  &  	38128.370 & 0.62  &     0.1116  &       5/2 &      \cite{Kramida2019} \\
	26126.907 &   2.94  &   0.147    &     5/2 &     \cite{Wickliffe1997}  &  	38433.920 & 14.9  &    1.043  &    5/2 &      \cite{Kramida2019} \\
	26439.491 &  0.806  &   0.05642  &     7/2 &     \cite{Wickliffe1997}  &  	38502.000 & 14.0  &    0.98  &       9/2 &      \cite{Kramida2019} \\
	26646.214 &   17.4  &   1.392    &     9/2 &     \cite{Wickliffe1997}  &  	38696.790 & 3.5  &     0.63  &       5/2 &      \cite{Kramida2019} \\
	26701.325 &   99.0  &   4.95     &     7/2 &     \cite{Wickliffe1997}  &  	39161.450 & 36.8  &    2.576  &    7/2 &      \cite{Kramida2019} \\
	26889.125 &    144  &   7.2      &     9/2 &     \cite{Wickliffe1997}  &     39206.840 & 2.7  &   0.486  &       9/2 &      \cite{Kramida2019} \\
		28024.010 &   3.80  &   0.19     &     9/2 &     \cite{Wickliffe1997}  &     39547.310 & 6.4   &  1.152  &      5/2   &  \cite{Kramida2019} \\
		28051.370 &   8.99  &   0.4495   &     5/2 &     \cite{Wickliffe1997}  &     39560.410 &  14.7  &    1.029  &    7/2  &   \cite{Kramida2019} \\
	28448.585 &   1.42  &   0.071    &     5/2 &     \cite{Wickliffe1997}  &     39768.790 &   5.29  &   0.2645   &     9/2 &     \cite{Wickliffe1997}  \\
	28555.799 &  0.668  &   0.04008  &     7/2 &     \cite{Wickliffe1997}  &               & &      && 
\end{tabular}
\end{table*}

\clearpage

\begin{table}
\caption{Correlation coefficients $r(C_k^{(i)},C_l^{(j)})$ among
four of the seven dispersion coefficients $C_k^{(i)}$ of Er$_2$ and
Tm$_2$. Their values and uncertainties are given in Table I of the
main text.  Coefficients are labeled by rank $k$ and index $i$
following the notation in the main text.  Correlation coefficients
with the remaining three dispersion coefficients follow from the
exact algebraic relationships among the dispersion coefficients.}
\label{tab:cc}
\begin{tabular}{c@{\ }|@{\ }rrrr}
\hline
\multicolumn{5}{c}{Homonuclear Erbium dimer}\\
\hline
 & \multicolumn{4}{c}{Correlation coeff.}\\
	$k,i\backslash l,j$   &      \multicolumn{1}{c}{0,1} & \multicolumn{1}{c}{2,1} & \multicolumn{1}{c}{0,2} & \multicolumn{1}{c}{0,3}\\
\hline
	0,1 & 1.00    & $-$0.38  & $-$0.50  & 0.32 \\
	2,1 & $-$0.38 & 1.00  & 0.37  & $-$1.00 \\
	0,2 & $-$0.50 &  0.37 &  1.00 &  $-$0.34 \\
	0,3 & 0.32    & $-$1.00  & $-$0.34 &  1.00\\
\hline
	\multicolumn{5}{c}{}\\
\multicolumn{5}{c}{Homonuclear Thulium dimer}\\
\hline
 & \multicolumn{4}{c}{Correlation coeff.}\\
	$k,i\backslash l,j$   &      \multicolumn{1}{c}{0,1} & \multicolumn{1}{c}{2,1} & \multicolumn{1}{c}{0,2} & \multicolumn{1}{c}{0,3}\\
\hline
0,1 & 1.00 & $-$0.03 & 0.15 & 0.05 \\
2,1 & $-$0.03 & 1.00 & $-$0.10 & $-$1.00 \\
0,2 & 0.15 & $-$0.10 & 1.00 & 0.09 \\
0,3 & 0.05 & $-$1.00 & 0.09 & 1.00  \\
\hline
\end{tabular}
\end{table}

\bibliography{References}

\end{document}